\RequirePackage{fix-cm}
\documentclass[smallextended]{svjour3}       
\smartqed  
\usepackage{graphicx,amssymb,amsfonts,mathrsfs,latexsym,amsmath,amssymb}
\usepackage{color}
\usepackage[all]{xy}
\journalname{}
\begin{document}

\title{On self-duality and hulls of cyclic codes over $\frac{\mathbb{F}_{2^m}[u]}{\langle u^k\rangle}$ with oddly even length
} \subtitle{}

\titlerunning{On self-duality and hulls of cyclic codes over $\frac{\mathbb{F}_{2^m}[u]}{\langle u^k\rangle}$ with oddly even length}

\author{Yonglin Cao$^1$ $\cdot$ Yuan Cao$^{1,2,3, \ast}$ $\cdot$ Fang-Wei Fu$^4$ 
}


\institute{Yuan Cao$^{\ast}$ (Corresponding author)\at
             \email{yuancao@sdut.edu.cn}\\
          Yonglin  Cao \at
             \email{ylcao@sdut.edu.cn}\\
           Fang-Wei Fu\at
             \email{fwfu@nankai.edu.cn} \\
          $^1$School of Mathematics and  Statistics, Shandong University of Technology, Zibo, Shandong 255091, China \\
          $^2$ Hunan Provincial Key Laboratory of Mathematical Modeling and Analysis in Engineering, Changsha University of Science and Technology,
          Changsha, Hunan 410114, China \\
          $^3$Hubei Key Laboratory of Applied Mathematics, Faculty of Mathematics and Statistics, Hubei University, Wuhan 430062, China \\
          $^4$Chern Institute of Mathematics and LPMC, and Tianjin Key Laboratory of Network and Data Security Technology, Nankai University, Tianjin 300071, China
          }

\date{Received: date / Accepted: date}

\maketitle

\begin{abstract}
Let $\mathbb{F}_{2^m}$ be a finite field of $2^m$ elements, and
$R=\mathbb{F}_{2^m}[u]/\langle u^k\rangle=\mathbb{F}_{2^m}+u\mathbb{F}_{2^m}+\ldots+u^{k-1}\mathbb{F}_{2^m}$ ($u^k=0$)
where $k$ is an integer satisfying $k\geq 2$.
For any odd positive integer $n$, an explicit representation for every self-dual cyclic code over $R$ of length $2n$
and a mass formula to count the number of these codes are given first. Then
a generator matrix  is provided for the
self-dual and $2$-quasi-cyclic code of length $4n$ over $\mathbb{F}_{2^m}$ derived by every self-dual cyclic code
of length $2n$ over $\mathbb{F}_{2^m}+u\mathbb{F}_{2^m}$ and a Gray map from $\mathbb{F}_{2^m}+u\mathbb{F}_{2^m}$
onto $\mathbb{F}_{2^m}^2$. Finally, the hull of
each cyclic code with length $2n$ over $\mathbb{F}_{2^m}+u\mathbb{F}_{2^m}$ is determined and
all distinct self-orthogonal cyclic codes of length $2n$ over $\mathbb{F}_{2^m}+u\mathbb{F}_{2^m}$ are
listed.

\end{abstract}

\section{Introduction and preliminaries}
\label{intro}
The class of self-dual codes is an interesting topic in coding theory duo to
their connections to other fields of mathematics such as Lattices, Cryptography, Invariant Theory, Block designs [5], etc.
A common theme for the construction of self-dual codes is the use of a computer search. In order to make this search feasible, special construction methods have been used to reduce the search field.
In many instances, self-dual codes have been found by first finding a code over a ring and then mapping
this code onto a code over a subring through a map that preserves duality. In the literatures, the mappings typically map
to codes over $\mathbb{F}_2$, $\mathbb{F}_4$ and $\mathbb{Z}_4$ since codes over these rings have had the most use.

\par
   Throughout this paper, let $\mathbb{F}_{2^m}$ be a finite field of $2^m$ elements and denote
\begin{center}
$R=\frac{\mathbb{F}_{2^m}[u]}{\langle u^k\rangle}=\mathbb{F}_{2^m}
+u\mathbb{F}_{2^m}+\ldots+u^{k-1}\mathbb{F}_{2^m} \ (u^k=0),$
\end{center}
 where $k\in \mathbb{Z}^{+}$ satisfying $k\geq 2$.
Then $R$ is a finite chain ring. Let $N$ be a fixed positive integer and
$R^N=\{(a_0,a_1,\ldots,a_{N-1})\mid a_0,a_1,\ldots,a_{N-1}\in R\}$
be an $R$-free module with the usual componentwise addition and scalar multiplication by elements of $R$. Then an
$R$-submodule $\mathcal{C}$ of $R^N$ is called a \textit{linear code} over
$R$ of length $N$. Moreover, the linear code $\mathcal{C}$ is said to be \textit{cyclic} if
$(c_{N-1},c_0,\ldots,c_{N-2})\in \mathcal{C}$ for all $(c_0,\ldots,c_{N-2},c_{N-1})\in \mathcal{C}$. The usual Euclidean inner
product on $R^N$ is defined $[\alpha,\beta]=\sum_{i=0}^{N-1}a_ib_i\in R$ for
all $\alpha=(a_0,a_1,\ldots,a_{N-1}), \beta=(b_0,b_1,\ldots,b_{N-1})\in R^N$. Then
the \textit{dual code} of $\mathcal{C}$ is defined by
$\mathcal{C}^{\bot}=\{\beta\in R^N\mid [\alpha,\beta]=0, \ \forall \alpha\in \mathcal{C}\}$, and $\mathcal{C}$ is said to be \textit{self-dual} (resp. \textit{self-orthogonal}) if $\mathcal{C}=\mathcal{C}^{\bot}$ (resp. $\mathcal{C}\subseteq \mathcal{C}^{\bot}$).

\par
  When $k=2$ and $m=1$, it is well known [6, 10] that the Gray map from
$\mathbb{F}_2+u\mathbb{F}_2$ onto $\mathbb{F}_2^2$ defined by $\phi: a+bu\mapsto (b,a+b)$
for all $a,b\in \mathbb{F}_2$ is an $\mathbb{F}_2$-linear space isomorphism and the image
of a self-dual codes over $\mathbb{F}_2+u\mathbb{F}_2$ under $\phi$ is always a self-dual code over $\mathbb{F}_2$.
Ling and Sol\'{e} studied Type II codes over the ring $\mathbb{F}_4+u\mathbb{F}_4$ in [16], which was later generalized to the ring $\mathbb{F}_{2^m}+u\mathbb{F}_{2^m}$
in [4]. These rings behave similar to the oft-studied ring $\mathbb{F}_2+u\mathbb{F}_2$ in the literature. The common theme in the aforementioned
works is that a distance and duality preserving Gray map can be defined that takes codes over those rings to binary codes,
preserving the linearity, the weight distribution and the duality. There were many
literatures on the construction of binary self-dual codes from various kind of codes over $\mathbb{F}_{2^m}+u\mathbb{F}_{2^m}$
for $m=1,2$. For some of these constructions, we refer to [10], [12--16].

\par
  For any finite field $\mathbb{F}_q$ of $q$ elements, where $q$ is a power of some prime number,
several different Gray type maps were defined, in similar fashion
obtaining different notions of distance for linear codes over $\frac{\mathbb{F}_q[u]}{\langle u^k\rangle}$, and
a method to obtain explicitly
new self-dual codes of larger length was presented from self-dual code over $\frac{\mathbb{F}_q[u]}{\langle u^k\rangle}$ in
[3]. Hence the construction and enumeration for self-dual codes over $\frac{\mathbb{F}_q[u]}{\langle u^k\rangle}$ for various
prime power $q$ and positive integer $k$ becomes a central topic in coding theory over finite rings.

\par
  Let $\frac{R[x]}{\langle x^N-1\rangle}=\{\sum_{i=0}^{N-1}a_ix^i\mid a_0,a_1,\ldots,a_{N-1}\in R\}$
in which the arithmetic is done modulo $x^N-1$.
In this paper, cyclic codes over
$R$ of length $N$ are identified with ideals of the ring $\frac{R[x]}{\langle x^N-1\rangle}$, under the
identification map
$$\sigma: R^N \rightarrow \frac{R[x]}{\langle x^N-1\rangle} \ {\rm via} \ \sigma: (a_0,a_1,\ldots,a_{N-1})\mapsto
\sum_{i=0}^{N-1}a_ix^i$$
($\forall a_0,a_1,\ldots,a_{N-1}\in R$). Moreover, ideals of $\frac{R[x]}{\langle x^N-1\rangle}$
are called \textit{simple-root cyclic codes} over $R$ when $N$ is relatively prime to the characteristic of $R$ and
called \textit{repeated-root cyclic codes} otherwise.

\par
   There were many literatures on cyclic codes of length $N$ over finite chain rings $R=\mathbb{F}_{2^m}[u]/\langle u^k\rangle$ for various positive integers $m, k, N$. For example: When $k=2$, Bonnecaze and Udaya [6] investigated cyclic codes and self-dual codes over $\mathbb{F}_2+u\mathbb{F}_2$ for odd length $N$. Norton and S\u{a}l\u{a}gean [17] discussed simple-root cyclic codes over an arbitrary finite chain ring $R$ systematically.
    Dinh [9] studied constacyclic codes and constacyclic codes over
Galois extension rings of $\mathbb{F}_2+u\mathbb{F}_2$ of length $N=2^s$.

\par
   When $k\geq 3$, in 2007 Abualrub and Siap [1] studied cyclic codes over the rings $\mathbb{Z}_2+u\mathbb{Z}_2$ and $\mathbb{Z}_2+u\mathbb{Z}_2+u^2\mathbb{Z}_2$ of the length $n$,
where either $n$ is odd or $n=2k$ ($k$ is odd) or $n$ is a power of $2$. This paper did not investigate self-dual
cyclic codes over ring $\mathbb{Z}_2+u\mathbb{Z}_2$ and $\mathbb{Z}_2+u\mathbb{Z}_2+u^2\mathbb{Z}_2$, but
asked a question:

\noindent
$\diamondsuit$ \textsf{Open problems include the study of self-dual codes and their
properties}.

\par
  In 2011, Al-Ashker and Hamoudeh [2] extended some of the results in [1],
and studied cyclic codes of an arbitrary length over
the ring $Z_2+uZ_2+u^2Z_2+\ldots+u^{k-1}Z_2$ with $u^k=0$. The rank and minimal spanning set of this family of codes were studied and two open problems were
asked:

\noindent
 $\diamondsuit$ \textsf{The study of cyclic codes of an arbitrary length over $Z_p+uZ_p+u^2Z_p+\ldots+u^{k-1}Z_p$,
where $p$ is a prime integer, $u^k=0$, and the study of dual and self-dual codes and their properties over these rings}.

\par
  In 2015, Singh et al. [19] studied cyclic
code over the ring $R_{k,p}=\mathbb{Z}_p[u]/\langle u^k\rangle$ $=Z_p+uZ_p+u^2Z_p+\ldots+u^{k-1}Z_p$ for any prime integer $p$ and positive integer $N$. If $n$ is not relatively prime to $p$, every cyclic code $C_k$ over $R_{k,p}$ of length $n$ is given by
[19, Theorem 3.3]:

\begin{enumerate}

\item
  \textit{$C_k=\langle g(x)+up_1(x) + u^2 p_2(x) + \ldots + u^{k-1} p_{k-1}(x)\rangle$ where $g(x)$ and $p_i(x)$ are
polynomials in $\mathbb{Z}_p[x]$ with $g(x)|(x^n-1)$ mod $p$, $(g(x)+up_1(x) + u^2 p_2(x) + \ldots + u^{k-1} p_{i-1}(x))\mid
(x^n-1)$ in $R_{i,p}$ and ${\rm deg}p_i < {\rm deg}p_{i-1}$ for all $1<i\leq k$}. Or

\item
 \textit{$C_k=\langle g(x)+up_1(x) + u^2 p_2(x) + \ldots + u^{k-1} p_{k-1}(x),
 ua_1(x) + u^2q_1(x) +\ldots+
u^{k-1}q_{k-2}(x),u^2a_2(x) + u^3l_1(x)+\ldots+u^{k-1}l_{k-3}(x),\ldots,
u^{k-2}a_{k-2}(x) + u^{k-1}t_1(x),u^{k-1}a_{k-1}(x)\rangle$ with $a_{k-1}(x)| a_{k-2}(x)|\ldots|
a_2(x)| a_1(x)| (x^n-1)$ mod $p$, $a_{k-2}(x)|p_1(x)\cdot\frac{x^n-1}{g(x)}$,$\ldots$, $a_{k-1}(x)|t_1(x)\cdot\frac{x^n-1}{a_{k-2}(x)}$,
$\ldots$, $a_{k-1}(x)|p_{k-1}(x)\cdot \frac{x^n-1}{g(x)}\cdots\frac{x^n-1}{a_{k-2}(x)}$. Moreover,
${\rm deg}p_{k-1}<{\rm deg}a_{k-1}, \ldots, {\rm deg}t_1<{\rm deg}a_{k-1},\ldots$, and ${\rm deg}p_1<
{\rm deg}a_{k-2}$}.

\end{enumerate}

\noindent
Furthermore, a set of
generators, the rank and the Hamming distance of these codes were investigated, but \textsf{the dual code and
self-duality for each cyclic code over $R_{k,p}$ were not considered} in [19].

\par
   In 2016, Chen et al. [8] given some new necessary and sufficient conditions for the existence of nontrivial self-dual simple-root cyclic codes over finite commutative chain rings and studied explicit enumeration formulas for these codes. But
\textsf{self-dual repeated-root cyclic codes over finite commutative chain rings were not considered} in [8].

\par
  In 2015, Sangwisut et al. [18] studied the hulls of cyclic and negacyclic codes over a finite field $\mathbb{F}_q$.
Based on the characterization of their generator polynomials, the dimensions of the hulls of cyclic and negacyclic codes over $\mathbb{F}_q$ were determined and the enumerations for hulls of cyclic codes and negacyclic codes over $\mathbb{F}_q$ were established.
However, in the literature, the representation for the hulls of repeated-root cyclic codes over the ring
$\mathbb{F}_q+u\mathbb{F}_q$ ($u^2=0$) have not been well studied.

\par
  In 2016, we [7] given a different approach from [1], [2] and [19] to study
cyclic code over $R=\frac{\mathbb{F}_{2^m}[u]}{\langle u^k\rangle}$ of length $2n$
for any odd positive integer $n$. We provided an explicit representation for each cyclic
code, given clear formulas to calculate
the number of codewords in each code and the number of all these cyclic codes respectively. Precisely, we determined the dual code for each code and presented a criterion to judge whether
a cyclic code over $R$ of length $2n$ is self-dual. Based on that, in this paper we study
the following questions:
\begin{itemize}
\item
 Give a clear expression for the Mass formula to count the number of all distinct
self-dual and cyclic codes over $R$ of length $2n$, using only the odd positive integer $n$,
the number $2^m$ of elements in $\mathbb{F}_{2^m}$ and integer $k\geq 2$.

\vskip 2mm\item
 Provide an explicit representation for all distinct
self-dual cyclic codes over $R$ of length $2n$ for any $k\geq 2$.

\vskip 2mm\item
  Give a generator matrix precisely for the
self-dual and $2$-quasi-cyclic code of length $4n$ over $\mathbb{F}_{2^m}$ derived by every self-dual cyclic code
of length $2n$ over $\mathbb{F}_{2^m}+u\mathbb{F}_{2^m}$ and a Gray map $\phi$ from $\mathbb{F}_{2^m}+u\mathbb{F}_{2^m}$
onto $\mathbb{F}_{2^m}^2$.

\vskip 2mm\item
  Determine the hull of
each cyclic code with length $2n$ over $\mathbb{F}_{2^m}+u\mathbb{F}_{2^m}$ and
list all distinct self-orthogonal cyclic code of length $2n$ over $\mathbb{F}_{2^m}+u\mathbb{F}_{2^m}$.
\end{itemize}

\par
   The present paper is organized as follows.
In Section 2, we introduce necessary notations and sketch known results for cyclic codes over $R$ of length $2n$
needed in this paper. In Section 3, we give an explicit representation and enumeration for self-dual cyclic codes over $R$ of length $2n$. Precisely, we provide a clear Mass formula for the number of all these codes.
In Section 4, we determine the generator matrix for each
self-dual and $2$-quasi-cyclic code of length $4n$ over $\mathbb{F}_{2^m}$ derived by a self-deal cyclic code
of length $2n$ over $\mathbb{F}_{2^m}+u\mathbb{F}_{2^m}$.
As an application, we list all $945$ self-dual cyclic codes of length $30$ over $\mathbb{F}_2+u\mathbb{F}_2$. In Section 5, we determine the hull
of each cyclic code
of length $2n$ over $\mathbb{F}_{2^m}+u\mathbb{F}_{2^m}$, and
give an explicit representation and enumeration for self-orthogonal cyclic codes of length $2n$
 over $\mathbb{F}_{2^m}+u\mathbb{F}_{2^m}$.
Section 6 concludes the paper.

\section{Known results for cyclic codes over $R$ of length $2n$}
In this section, we list the necessary notations and known results for cyclic codes  of length $2n$ over the ring $R=\frac{\mathbb{F}_{2^m}[u]}{\langle u^k\rangle}$
needed in the following sections.

\par
   As $n$ is odd, there are pairwise coprime monic irreducible polynomials $f_1(x)=x-1,f_2(x),\ldots,f_r(x)$ in $\mathbb{F}_{2^m}[x]$
such that
\begin{equation}
x^n-1=f_1(x)f_2(x)\ldots f_r(x).
\end{equation}
Then we have
$x^{2n}-1=(x^n-1)^2=f_1(x)^2\ldots f_r(x)^2$.

\par
Let $1\leq j\leq r$. We assume ${\rm deg}(f_j(x))=d_j$ and denote $F_j(x)=\frac{x^{2n}-1}{f_j(x)^2}$. Then ${\rm gcd}(F_j(x),f_j(x)^2)=1$
and there exist $a_j(x),b_j(x)\in\mathbb{F}_{2^m}[x]$ such that
$$
a_j(x)F_j(x)+b_j(x)f_j(x)^2=1.
$$
  As in [7], we will adopt the following notations in this paper:

\begin{description}
\item{$\bullet$}
  $\mathcal{A}=\frac{\mathbb{F}_{2^m}[x]}{\langle x^{2n}-1\rangle}=\{\sum_{i=0}^{2n-1}a_ix^i\mid
a_i\in \mathbb{F}_{2^m}, \ i=0,1,\ldots, 2n-1\}$ in which the arithmetic is done modulo $x^{2n}-1$.

\vskip 2mm\item{$\bullet$}
  Let $\varepsilon_j(x)\in \mathcal{A}$ be defined by
$$\varepsilon_j(x)\equiv a_j(x)F_j(x)=1-b_j(x)f_j(x)^2 \ ({\rm mod} \ x^{2n}-1).$$
Then $\varepsilon_j(x)^2=\varepsilon_j(x)$ and $\varepsilon_j(x)\varepsilon_l(x)=0$ in the ring
$\mathcal{A}$ for all $j\neq l$ and $j,l=1,\ldots,r$ (cf. [7] Theorem 2.3).

\vskip 2mm\item{$\bullet$}
  $\mathcal{K}_j=\frac{\mathbb{F}_{2^m}[x]}{\langle f_j(x)^2\rangle}=\{\sum_{i=0}^{2d_j-1}a_ix^i\mid
a_i\in \mathbb{F}_{2^m}, \ i=0,1,\ldots, 2d_j-1\}$ in which the arithmetic is done modulo $f_j(x)^2$.

\vskip 2mm\item{$\bullet$}
  $\mathcal{F}_j=\frac{\mathbb{F}_{2^m}[x]}{\langle f_j(x)\rangle}=\{\sum_{i=0}^{d_j-1}a_ix^i\mid
a_i\in \mathbb{F}_{2^m}, \ i=0,1,\ldots, d_j-1\}$ in which the arithmetic is done modulo $f_j(x)$. Then
${\cal F}_j$ is an extension field of $\mathbb{F}_{2^m}$ with $2^{md_j}$ elements.

\vskip 2mm\item{$\bullet$}
  For each $\Upsilon\in \{\mathcal{A},\mathcal{K}_j,\mathcal{F}_j\}$, we set
$$
\frac{\Upsilon[u]}{\langle u^k\rangle}
  =\{\alpha_0+u\alpha_1+\ldots+u^{k-1}\alpha_{k-1}\mid \alpha_0,\alpha_1,\ldots,\alpha_{k-1}\in \Upsilon\} \ (u^k=0).
$$

\end{description}

\noindent
  \textit{Remark} $\mathcal{F}_j$ is a finite field with
operations defined by the usual polynomial operations modulo $f_j(x)$, $\mathcal{K}_j$ is a finite chain ring with
operations defined by the usual polynomial operations modulo $f_j(x)^2$ (cf. Lemma 2.4(v) in [7])
and $\mathcal{A}$ is a finite principal ideal ring with
operations defined by the usual polynomial operations modulo $x^{2n}-1$. As in [7], we
adopt the following points of view:
$$\mathcal{F}_j\subseteq \mathcal{K}_j \subseteq \mathcal{A}
\ {and} \ \frac{\mathcal{F}_j[u]}{\langle u^k\rangle}\subseteq \frac{\mathcal{K}_j[u]}{\langle u^k\rangle}
\subseteq \frac{\mathcal{A}[u]}{\langle u^k\rangle}$$
only as sets. Obviously, $\mathcal{F}_j$ is not a subfield
of $\mathcal{K}_j$, $\mathcal{K}_j$ is not a subring of $\mathcal{A}$ when $n\geq 2$;
$\frac{\mathcal{F}_j[u]}{\langle u^k\rangle}$ is not a subring
of $\frac{\mathcal{K}_j[u]}{\langle u^k\rangle}$ and  $\frac{\mathcal{K}_j[u]}{\langle u^k\rangle}$ is not a subring of $\frac{\mathcal{A}[u]}{\langle u^k\rangle}$ when $n\geq 2$.

\vskip 3mm
\par
   For any $\alpha(x)=\sum_{i=0}^{2n-1}\alpha_ix^i\in \frac{R[x]}{\langle x^{2n}-1\rangle}$
where $\alpha_i=\sum_{j=0}^{k-1}a_{i,j}u^j\in R$ with $a_{i,j}\in \mathbb{F}_{2^m}$ for all
$i=0,1,\ldots,2n-1$ and $j=0,1,\ldots,k-1$, we define
$$\Psi(\alpha(x))=a_0(x)+a_1(x)u+\ldots+a_{k-1}(x)u^{k-1}\in \frac{{\cal A}[u]}{\langle u^k\rangle}$$
where
$a_j(x)=\sum_{i=0}^{2n-1}a_{i,j}x^i\in {\cal A}$ for all $j=0,1,\ldots,k-1.$ Then
the map $\Psi$ is a ring isomorphism from $\frac{R[x]}{\langle x^{2n}-1\rangle}$ onto
$\frac{{\cal A}[u]}{\langle u^k\rangle}$ (cf. [7] Lemma 2.2).

\par
  As in [7], we will identify $\frac{R[x]}{\langle x^{2n}-1\rangle}$ with $\frac{{\cal A}[u]}{\langle u^k\rangle}$
under this ring isomorphism $\Psi$ in the rest of this paper. From this, we deduce that
$\mathcal{C}$ is a cyclic code over $R$ of length $2n$, i.e. $\mathcal{C}$ is an
ideal of $\frac{R[x]}{\langle x^{2n}-1\rangle}$, if and only if $\mathcal{C}$ is an
ideal of the ring $\frac{\mathcal{A}[u]}{\langle u^k\rangle}$. Then in order to determine cyclic codes over $R$ of length $2n$,
it is sufficient to determine ideals of the ring $\frac{{\cal A}[u]}{\langle u^k\rangle}$.

\par
  First, every ideal of the ring $\frac{{\cal A}[u]}{\langle u^k\rangle}$ can be determined by
a unique ideal of $\frac{{\cal K}_j[u]}{\langle u^k\rangle}$ for each $j=1,\ldots,r$.
See the following lemma.

\vskip 3mm
\noindent
  {\bf Lemma 2.1} (cf. [7] Theorem 2.3) \textit{Let $\mathcal{C}\subseteq \frac{\mathcal{A}[u]}{\langle u^k\rangle}$. Then
$\mathcal{C}$ is a cyclic code over $R$ of length $2n$ if and only if for each integer $j$, $1\leq j\leq r$, there is a unique
ideal $C_j$ of the ring $\frac{\mathcal{K}_j[u]}{\langle u^k\rangle}$ such that}
$$\mathcal{C}=\bigoplus_{j=1}^r\varepsilon_j(x)C_j=\sum_{j=1}^r\varepsilon_j(x)C_j
\ ({\rm mod} \ x^{2n}-1),$$
\textit{where $\varepsilon_j(x)C_j=\{\varepsilon_j(x)c_j(x) \ ({\rm mod} \ x^{2n}-1)\mid c_j(x)\in C_j\}\subseteq
\frac{\mathcal{A}[u]}{\langle u^k\rangle}$ for all $j=1,\ldots,r$. Moreover, the number of codewords in $\mathcal{C}$ is
equal to $\prod_{j=1}^r|C_j|$}.

\vskip 3mm\par
   Then order to present all distinct ideals of the ring ${\cal K}_j[u]/\langle u^k\rangle$ for all $j=1,\ldots,r$,
we need the following lemma.

\vskip 3mm\noindent
  {\bf Lemma 2.2} (cf. [7] Lemma 2.4 (ii)--(iv)) \textit{Using the notations above, for any $1\leq j\leq r$ and $1\leq s\leq k$ we have the following}:

\begin{description}
\vskip 2mm
\item{(i)}
\textit{The ring $\frac{{\cal F}_j[u]}{\langle u^s\rangle}$
is a finite commutative chain ring with the unique maximal ideal $u(\frac{{\cal F}_j[u]}{\langle u^s\rangle})$,
the nilpotency index of $u$ is equal to $s$ and the residue field of $\frac{{\cal F}_j[u]}{\langle u^s\rangle}$ is $(\frac{{\cal F}_j[u]}{\langle u^s\rangle})/u(\frac{{\cal F}_j[u]}{\langle u^s\rangle})
\cong {\cal F}_j$}.

\vskip 2mm
\item{(ii)}
 \textit{Every element $\alpha$ of $\frac{{\cal F}_j[u]}{\langle u^s\rangle}$
has a unique $u$-adic expansion}:
$$\alpha=b_0(x)+ub_1(x)+\ldots+u^{s-1}b_{s-1}(x), \ b_0(x),b_1(x),\ldots,b_{s-1}(x)\in {\cal F}_j$$
\textit{Moreover, $\alpha$ is an invertible element of $\frac{{\cal F}_j[u]}{\langle u^s\rangle}$ if and only if
$b_0(x)\neq 0$. The set of all invertible elements of $\frac{{\cal F}_j[u]}{\langle u^s\rangle}$ is denoted
by $(\frac{{\cal F}_j[u]}{\langle u^s\rangle})^\times$}.

\vskip 2mm
\item{(iii)}
 \textit{$|(\frac{{\cal F}_j[u]}{\langle u^s\rangle})^{\times}|=(2^{md_j}-1)2^{(s-1)md_j}$}.
\end{description}

\vskip 3mm
\par
  Now, using the notation of Lemma 2.2 all ideals of $\frac{{\cal K}_j[u]}{\langle u^k\rangle}$ are listed by the following
lemma.

\vskip 3mm
\noindent
  {\bf Lemma 2.3} ([7] Theorem 2.6) \textit{Let $1\leq j\leq r$. Then all distinct ideals of the ring
  $\frac{{\cal K}_j[u]}{\langle u^k\rangle}$ are
given by the following table}:
\begin{center}
\begin{tabular}{lll}\hline
  number of ideals  &  $C_j$ (ideal of $\frac{{\cal K}_j[u]}{\langle u^k\rangle}$)    &   $|C_j|$  \\ \hline
 $k+1$  & $\bullet$ $\langle u^i\rangle$ \ $(0\leq i\leq k)$ & $2^{2md_j(k-i)}$ \\
 $k$     & $\bullet$  $\langle u^s f_j(x)\rangle$ \ $(0\leq s\leq k-1)$ &  $2^{md_j(k-s)}$  \\
 $\Omega_1(2^{md_j},k)$ & $\bullet$   $\langle u^i+u^tf_j(x)\omega\rangle$ &  $2^{2md_j(k-i)}$ \\
                                 & \ \ ($\omega\in (\frac{{\cal F}_j[u]}{\langle u^{i-t}\rangle})^{\times}$, $t\geq 2i-k$, & \\
                                 & \ \  $ 0\leq t<i\leq k-1)$                                 &              \\
 $\Omega_2(2^{md_j},k)$     & $\bullet$   $\langle u^i+u^tf_j(x)\omega \rangle$ &  $2^{md_j(k-t)}$ \\
                                 & \ \  ($\omega\in (\frac{{\cal F}_j[u]}{\langle u^{k-i}\rangle})^{\times}$, $t< 2i-k$, & \\
                                 & \ \  $ 0\leq t<i\leq k-1)$        \\
  $\frac{1}{2}k(k-1)$    & $\bullet$    $\langle u^i,u^sf_j(x)\rangle$ &  $2^{md_j(2k-(i+s))}$ \\
                              &  $ \ \ (0\leq s<i\leq k-1)$ & \\
  $(2^{md_j}-1)$    &  $\bullet$   $\langle u^i+u^tf_j(x)\omega, u^sf_j(x)\rangle$ &  $2^{md_j(2k-(i+s))}$ \\
      $\cdot \Gamma(2^{md_j},k)$ & \ \  $(\omega\in (\frac{{\cal F}_j[u]}{\langle u^{s-t}\rangle})^{\times}$,  $i+s\leq k+t-1$, & \\
                              & \ \ $0\leq t<s<i\leq k-1)$ & \\ \hline
\end{tabular}
\end{center}

\vskip 2mm\noindent
\textit{where $|C_j|$ is the number of elements in $C_j$, and}

\begin{description}
\vskip 2mm
\item{$\diamond$}
   $\Omega_1(2^{md_j},k)=\left\{\begin{array}{ll}\frac{2^{md_j(\frac{k}{2}+1)}+
2^{md_j\cdot\frac{k}{2}}-2}{2^{md_j}-1}-(k+1), & {\rm if} \ k \ {\rm is} \ {\rm even};\cr
\frac{2(2^{md_j\cdot\frac{k+1}{2}}-1)}{2^{md_j}-1}-(k+1), & {\rm if} \ k \ {\rm is} \ {\rm odd}.\end{array}\right.$

\vskip 2mm
\item{$\diamond$}
  $\Omega_2(2^{md_j},k)=\left\{\begin{array}{ll}(2^{md_j}-1)\sum_{i=\frac{k}{2}+1}^{k-1}(2i-k)2^{md_j(k-i-1)}, & {\rm if} \ k \ {\rm is} \ {\rm even};\cr
(2^{md_j}-1)\sum_{i=\frac{k+1}{2}}^{k-1}(2i-k)2^{md_j(k-i-1)}, & {\rm if} \ k \ {\rm is} \ {\rm odd}.\end{array}\right.$

\vskip 2mm
\item{$\diamond$}
 \textit{$\Gamma(2^{md_j},k)$ can be calculated by the following recurrence formula}:

\begin{description}
\vskip 2mm
\item{}
   \textit{$\Gamma(2^{md_j},\rho)=0$ for $\rho=1,2,3$, $\Gamma(2^{md_j},\rho)=1$ for $\rho=4$};

\vskip 2mm
\item{}
  \textit{$\Gamma(2^{md_j},\rho)=\Gamma(2^{md_j},\rho-1)+\sum_{s=1}^{\lfloor\frac{\rho}{2}\rfloor-1}(\rho-2s-1)2^{md_j(s-1)}$ for $\rho\geq 5$}.
\end{description}
\end{description}

\vskip 2mm \noindent
  \textit{Therefore, the number of all
distinct ideals of the ring  ${\cal K}_j[u]/\langle u^k\rangle$ is equal to}
\begin{center}
$N_{(2^m,d_j,k)}=1+\frac{k(k+3)}{2}+\Omega_1(2^{md_j},k)+\Omega_2(2^{md_j},k)+(2^{md_j}-1)\Gamma(2^{md_j},k)$.
\end{center}

\vskip 3mm
\par
  As the end of this section, we give an explicit formula to count the number
of all cyclic codes over $R$ of length $2n$.

\vskip 3mm \noindent
   {\bf Theorem 2.4} \textit{Using the notation above, let $1\leq j\leq r$.}

\vskip 2mm
\begin{description}
\item{(i)}
  \textit{The number of all
distinct ideals of the ring  ${\cal K}_j[u]/\langle u^k\rangle$ is }
\begin{equation}
N_{(2^m,d_j,k)}=\left\{\begin{array}{ll}\sum_{i=0}^{\frac{k}{2}}(1+4i)2^{(\frac{k}{2}-i)md_j}, & {\rm if} \ k \ {\rm is} \ {\rm even}; \cr \sum_{i=0}^{\frac{k-1}{2}}(3+4i)2^{(\frac{k-1}{2}-i)md_j}, & {\rm if} \ k \ {\rm is} \ {\rm odd}.\end{array}\right.
\end{equation}
\textit{Precisely, we have}
\begin{description}
\item{}
$N_{(2^m,d_j,k)}=\frac{(2^{md_j}+3)2^{(\frac{k}{2}+1)md_j}-2^{md_j}(2k+5)+2k+1}{(2^{md_j}-1)^2}$,
\textit{when $k$ is even};

\vskip 2mm\item{}
$N_{(2^m,d_j,k)}=\frac{(3\cdot 2^{md_j}+1)2^{(\frac{k-1}{2}+1)md_j}-2^{md_j}(2k+5)+2k+1}{(2^{md_j}-1)^2}$,
\textit{when $k$ is odd}.
\end{description}

\vskip 2mm
\item{(ii)}
  \textit{The number of cyclic codes over $\mathbb{F}_{2^m}[u]/\langle u^k\rangle$
of length $2n$ is equal to}
\begin{description}
\vskip 2mm\item{}
$\prod_{j=1}^r\frac{(2^{md_j}+3)2^{(\frac{k}{2}+1)md_j}-2^{md_j}(2k+5)+2k+1}{(2^{md_j}-1)^2}$,
\textit{when $k$ is even};

\vskip 2mm
\item{}
$\prod_{j=1}^r\frac{(3\cdot 2^{md_j}+1)2^{(\frac{k-1}{2}+1)md_j}-2^{md_j}(2k+5)+2k+1}{(2^{md_j}-1)^2}$,
\textit{when $k$ is odd}.
\end{description}
\end{description}

\vskip 3mm \noindent
   \textit{Proof} (i) By the mathematical induction on $k$, one can easily verify
that the equation (2) holds.

\par
  Now, let $k=2s+1$ where $s$ is a positive integer, and denote $q=2^{md_j}$. Then
we have $N_{(2^m,d_j,k)}=\sum_{i=0}^{s}(3+4i)q^{s-i}=3\sum_{i=0}^{s}q^{s-i}+4q^s\sum_{i=0}^{s}iq^{-i}$
in which $\sum_{i=0}^{s}q^{s-i}=\frac{q^{s+1}-1}{q-1}$. Then by
\begin{eqnarray*}
\sum_{i=1}^{s}ix^{i-1}&=&\frac{d}{dx}(\sum_{i=0}^{s}x^{i})=\frac{d}{dx}\left(\frac{x^{s+1}-1}{x-1}\right)\\
 &=& \frac{(s+1)x^{s}(x-1)-(x^{s+1}-1)}{(x-1)^2},
\end{eqnarray*}
we have
\begin{eqnarray*}
q^s\sum_{i=0}^{s}iq^{-i}&=&q^{s-1}\sum_{i=1}^{s}i(q^{-1})^{i-1}=\frac{(s+1)q^{-s}(q^{-1}-1)-(q^{-(s+1)}-1)}{(q^{-1}-1)^2} \\
  &=& q^{s-1}\cdot \frac{q^{-s+1}}{(q-1)^2}\left(q^{s+1}-q(s+1)+s\right).
\end{eqnarray*}
From these, we deduce $N_{(2^m,d_j,k)}=3\cdot\frac{q^{s+1}-1}{q-1}+4\cdot\frac{1}{(q-1)^2}\left(q^{s+1}-q(s+1)+s\right)$, and hence
$N_{(2^m,d_j,k)}=\frac{(3q+1)q^{s+1}-q(4s+7)+4s+3}{(q-1)^2}$ where $s=\frac{k-1}{2}$.

\par
  When $k$ is even, the conclusion can be proved similarly. We omit this here.

\par
  (ii) It follows from (i) and Lemma 2.3.
\hfill $\Box$

\vskip 3mm\par
   For the special cases of $k=2,3,4,5$, we have the following conclusions.

\vskip 3mm \noindent
   {\bf Corollary 2.5} \textit{Let $2\leq k\leq 5$. The number of all
ideals in ${\cal K}_j[u]/\langle u^k\rangle$ is }
$$
N_{(2^m,d_j,k)}=\left\{\begin{array}{ll}5+2^{d_jm}, & {\rm when} \ k=2; \cr
 9+5\cdot 2^{d_jm}+2^{2d_jm}, & {\rm when} \ k=4; \cr
 7+3\cdot 2^{d_jm}, & {\rm when} \ k=3; \cr
 11+7\cdot 2^{d_jm}+3\cdot 2^{2d_jm}, & {\rm when} \ k=5.\end{array}\right.
$$

\vskip 3mm\par
   Finally, let $n=1$ and $m=1$. Then $r=1$ and $d_1=1$ in this case. We denote by
$L_k$ the number of ideals in the ring $\frac{(\mathbb{F}_2+u\mathbb{F}_2+\ldots+u^{k-1}\mathbb{F}_2)[x]}{\langle x^2-1\rangle}$, where
$k\geq 2$. By Theorem 2.4, we have that
$$L_k=N_{(2,1,k)}=\left\{\begin{array}{ll} 10\cdot 2^{\frac{k}{2}}-2k-9 & {\rm if} \ 2\mid k;
\cr 14\cdot 2^{\frac{k-1}{2}}-2k-9 & {\rm if} \ 2\nmid k.
  \end{array}\right.$$
For examples, we have $L_2=7$, $L_4=23$, $L_6=59$, $L_8=135$;
$L_3=13$, $L_5=37$, $L_7=89$ and $L_9=197$.

\section{An explicit representation and enumeration for self-dual cyclic codes over $R$ of length $2n$}
\noindent
   In this section, we give an explicit representation for self-dual cyclic codes over $R$ of length $2n$
and a precise mass formula to count the nuber of these codes.

For any polynomial $f(x)=\sum_{l=0}^da_lx^l\in \mathbb{F}_{2^m}[x]$ of degree $d\geq 1$, recall that
the \textit{reciprocal polynomial} of $f(x)$ is defined as $\widetilde{f}(x)=x^df(\frac{1}{x})=\sum_{l=0}^da_lx^{d-l}$, and
 $f(x)$ is said to be \textit{self-reciprocal} if $\widetilde{f}(x)=\delta f(x)$ for some $\delta\in \mathbb{F}_{2^m}^{\times}$. Then by Equation (1) in Section 2, it follows that
$$x^{n}-1=x^{n}+1=\widetilde{f}_1(x)\widetilde{f}_2(x)\ldots \widetilde{f}_r(x).$$
Since $f_1(x)=x+1,f_2(x),\ldots,f_r(x)$ are pairwise coprime monic irreducible polynomials in $\mathbb{F}_{2^m}[x]$,
 $\widetilde{f}_1(x)=x+1,\widetilde{f}_2(x),\ldots, \widetilde{f}_r(x)$  are pairwise coprime irreducible polynomials in $\mathbb{F}_{2^m}[x]$ as well. Hence for each integer $j$, $1\leq j\leq r$,
there is a unique integer $j^{\prime}$, $1\leq j^{\prime}\leq r$, such that
$$\widetilde{f}_j(x)=\delta_jf_{j^{\prime}}(x) \
{\rm where} \ \delta_j\in \mathbb{F}_{2^m}^{\times}.$$ Then
After a rearrangement of $f_2(x),\ldots,f_r(x)$, there are integers $\lambda,\epsilon$ such that

\vskip 2mm\par
  $\bullet$ $\lambda+2\epsilon=r$ where $\lambda\geq 1$ and $\epsilon\geq 0$;

\vskip 2mm\par
  $\bullet$ $\widetilde{f}_j(x)=\delta_jf_{j}(x)$, where $\delta_j\in \mathbb{F}_{2^m}^{\times}$,
for all $j=1,\ldots,\lambda$;

\vskip 2mm\par
  $\bullet$ $\widetilde{f}_j(x)=\delta_jf_{j+\epsilon}(x)$, where $\delta_j\in \mathbb{F}_{2^m}^{\times}$,
for all $j=\lambda+1,\ldots,\lambda+\epsilon$.

\vskip 2mm\par
  Let $1\leq j\leq r$. Since $f_{j}(x)^2$ is a divisor of $x^{2n}-1$, we have
$x^{2n}\equiv 1$ (mod $f_{j}(x)^2$), i.e. $x^{2n}=1$ in the ring $\mathcal{K}_j=\frac{\mathbb{F}_{2^m}[x]}{\langle f_{j}(x)^2\rangle}$.
This implies that
$$x^{-d}=x^{2n-d} \ {\rm in} \ \mathcal{K}_j[u]/\langle u^k\rangle, \ 1\leq d\leq 2n-1.$$
For any integer $s$, $1\leq s\leq k$, and $\omega=\omega(x)\in \frac{\mathcal{F}_j[u]}{\langle u^s\rangle}$,
by Lemma 2.2(ii) we know that $\omega(x)$ has a unique $u$-expansion:
$$\omega(x)=\sum_{i=0}^{s-1}u^{i}a_{i}(x), \
a_0(x),a_1(x),\ldots,a_{s-1}(x)\in \mathcal{F}_j.$$
To simplify the expressions, we adopt the following notation in this paper:

\begin{description}
\vskip 2mm
\item{$\bullet$}
   $\widehat{\omega}=\omega(x^{-1})=a_0(x^{-1})+ua_1(x^{-1})+\ldots+u^{s-1}a_{s-1}(x^{-1})$
(mod $f_j(x)$), when $1\leq j\leq \lambda$;

\vskip 2mm
\item{$\bullet$}
  $\widehat{\omega}=\omega(x^{-1})=a_0(x^{-1})+ua_1(x^{-1})+\ldots+u^{s-1}a_{s-1}(x^{-1})$
(mod $f_{j+\epsilon}(x)$), when $\lambda+1\leq j\leq \lambda+\epsilon$.

\vskip 2mm
\item{$\bullet$}
  $\Theta_{j,s}=\{\omega\in (\frac{\mathcal{F}_j[u]}{\langle u^s\rangle})^\times
  \mid \omega+\delta_jx^{2n-d_j}\widehat{\omega}\equiv 0 \ ({\rm mod} \ f_j(x))\}$, where
$1\leq j\leq \lambda$ and $1\leq s\leq k-1$.
\end{description}

\vskip 2mm\par
  For self-dual cyclic codes over $R$, using the notation above and by Theorem 3.6 in [7] we have the following conclusion.

\vskip 3mm
\noindent
  {\bf Theorem 3.1} \textit{Using the notations above, all
distinct self-dual cyclic
codes over the ring $R$ of length $2n$ are give by}:
$${\cal C}=\left(\oplus_{j=1}^\lambda \varepsilon_j(x)C_j\right)\oplus
\left(\oplus_{j=\lambda+1}^{\lambda+\epsilon}(\varepsilon_{j}(x)C_{j}\oplus\varepsilon_{j+\epsilon}(x)C_{j+\epsilon})\right),$$
\textit{where $C_j$ is an ideal of ${\cal K}_j[u]/\langle u^k\rangle$ determined by the following conditions}:

\begin{description}
\vskip 2mm
\item{(i)}
\textit{If $1\leq j\leq \lambda$, $C_j$ is determined by the following conditions}:

\begin{description}
\vskip 2mm
\item{($\dag$)}
\textit{When $k$ is even, $C_j$ is given by one of the following six cases}:

\begin{description}
\item{($\dag$-1)}
 \textit{$C_j=\langle u^{\frac{k}{2}}\rangle$}.

\item{($\dag$-2)}
\textit{$C_j=\langle f_j(x)\rangle$}.

\item{($\dag$-3)}
\textit{$C_j=\langle u^{\frac{k}{2}}+u^{t}f_j(x)\omega\rangle$, where
$\omega \in\Theta_{j,\frac{k}{2}-t}$ and $0\leq t\leq \frac{k}{2}-1$}.

\item{($\dag$-4)}
\textit{$C_j=\langle u^{i}+f_j(x)\omega\rangle$, where
$\omega\in\Theta_{j,k-i}$ and $\frac{k}{2}+1\leq i\leq k-1$}.

\item{($\dag$-5)}
\textit{$C_j=\langle u^i,u^{k-i}f_j(x)\rangle$, where
$\frac{k}{2}+1\leq i\leq k-1$}.

\item{($\dag$-6)}
\textit{$C_j=\langle u^{i}+u^{t}f_j(x)\omega, u^{k-i}f_j(x)\rangle$, where
$\omega\in \Theta_{j,k-i-t}$}, $1\leq t<k-i$ and $\frac{k}{2}+1\leq i\leq k-1$.
\end{description}

\vskip 2mm
\item{($\ddag$)}
 \textit{When $k$ is odd, $C_j$ is given by one of the following four cases}:
\begin{description}
\item{($\ddag$-1)}
  \textit{$C_j=\langle f_j(x)\rangle$}.

\item{($\ddag$-2)}
  \textit{$C_j=\langle u^{i}+f_j(x)\omega\rangle$, where $\omega\in \Theta_{j,k-i}$
and $\frac{k+1}{2}\leq i\leq k-1$}.

\item{($\ddag$-3)}
\textit{$C_j=\langle u^i,u^{k-i}f_j(x)\rangle$, where
$\frac{k+1}{2}\leq i\leq k-1$}.

\item{($\ddag$-4)}
\textit{$C_j=\langle u^{i}+u^{t}f_j(x)\omega, u^{k-i}f_j(x)\rangle$, where
$\omega\in \Theta_{j,k-i-t}$, $1\leq t<k-i$ and $\frac{k+1}{2}\leq i\leq k-1$}.
\end{description}
\end{description}

\vskip 2mm
\item{(ii)} \textit{If $\lambda+1\leq j\leq \lambda+\epsilon$, then $(C_j, C_{j+\epsilon})$ is given by
one of the $N_{(2^m,d_j,k)}$ pairs listed in the below table}:
\end{description}
\begin{center}
\begin{tabular}{ll}\hline
  $C_j$ (mod $f_j(x)^2$) &  $C_{j+\epsilon}$  (mod $f_{j+\epsilon}(x)^2$)\\ \hline
$\bullet$ $\langle u^i\rangle$ \ $(0\leq i\leq k)$ & $\diamond$ $\langle u^{k-i}\rangle$  \\
$\bullet$  $\langle u^sf_j(x)\rangle$ \ ($0\leq s\leq k-1$) & $\diamond$ $\langle u^{k-s},f_{j+\epsilon}(x)\rangle$ \\
$\bullet$ $\langle u^i+u^tf_j(x)\omega\rangle$ & $\diamond$ $\langle u^{k-i}+u^{k+t-2i}f_{j+\epsilon}(x)\omega^{\prime}\rangle$ \\
      \ \ \ ($\omega\in (\frac{{\cal F}_j[u]}{\langle u^{i-t}\rangle})^{\times}$, &  \ \ \  $\omega^{\prime}=\delta_jx^{2n-d_j}\widehat{\omega}$
       (mod $f_{j+\epsilon}(x)$)\\
      \ \ \  $t\geq 2i-k, 0\leq t<i\leq k-1$) & \\
$\bullet$ $\langle u^i+f_j(x)\omega\rangle$  & $\diamond$  $\langle u^i+f_{j+\epsilon}(x)\omega^{\prime}\rangle$ \\
     \ \ \  $(\omega\in (\frac{{\cal F}_j[u]}{\langle u^{k-i}\rangle})^{\times}$, &  \ \ \   $\omega^{\prime}=\delta_jx^{2n-d_j}\widehat{\omega}$
       (mod $f_{j+\epsilon}(x)$)\\
    \ \ \   $2i>k, 0<i\leq k-1$) & \\
$\bullet$ $\langle u^i+u^{t}f_j(x)\omega\rangle$ & $\diamond$ $\langle u^{i-t}+f_{j+\epsilon}(x)\omega^{\prime},u^{k-i}f_{j+\epsilon}(x)\rangle$  \\
     \ \ \  $(\omega\in (\frac{{\cal F}_j[u]}{\langle u^{k-i}\rangle})^{\times}$, &  \ \ \  $\omega^{\prime}=\delta_jx^{2n-d_j}\widehat{\omega}$
       (mod $f_{j+\epsilon}(x)$) \\
    \ \ \   $t<2i-k, 1\leq t<i\leq k-1$)     & \\
$\bullet$ $\langle u^{i},u^{s}f_j(x)\rangle$ & $\diamond$ $\langle u^{k-s},u^{k-i}f_{j+\epsilon}(x)\rangle$ \\
      \ \ \   $(0\leq s<i\leq k-1)$  & \\
$\bullet$ $\langle u^{i}+f_j(x)\omega, u^{s}f_j(x)\rangle$ & $\diamond$ $\langle u^{k-s}+u^{k-i-s}f_{j+\epsilon}(x)\omega^{\prime}\rangle$ \\
    \ \ \   $(\omega\in (\frac{{\cal F}_j[u]}{\langle u^s\rangle})^{\times}$, &  \ \ \  $\omega^{\prime}=\delta_jx^{2n-d_j}\widehat{\omega}$
       (mod $f_{j+\epsilon}(x)$) \\
    \ \ \   $i+s\leq k-1, 1\leq s<i\leq k-1$) & \\
$\bullet$ $\langle u^{i}+u^{t}f_j(x)\omega, u^{s}f_j(x)\rangle$ & $\diamond$ $\langle u^{k-s}+u^{k+t-i-s}f_{j+\epsilon}(x)\omega^{\prime},$ \\
  \ \ \   $(\omega\in (\frac{{\cal F}_j[u]}{\langle u^{s-t}\rangle})^{\times}$,  &  \ \ \   $u^{k-i}f_{j+\epsilon}(x)\rangle$ \\
   \ \ \   $i+s\leq k+t-1$,  & \ \ \ $\omega^{\prime}=\delta_jx^{2n-d_j}\widehat{\omega}$
      (mod $f_{j+\epsilon}(x)$) \\
  \ \ \   $1\leq t<s<i\leq k-1)$ & \\ \hline
\end{tabular}
\end{center}

\vskip 3mm\par
   Hence in order to listed all self-dual cyclic codes
over $R$ of length $2n$, by Lemma 3.1 we need to determine
the set $\Theta_{j,s}$ of elements $\omega\in ({\cal F}_j[u]/\langle u^{s}\rangle)^{\times}$ satisfying
\begin{equation}
\omega+\delta_jx^{2n-d_j}\widehat{\omega}\equiv 0 \ ({\rm mod} \ f_j(x))
\end{equation}
for some integer $s$, $1\leq s\leq k-1$, and for all $j=1,\ldots,\lambda$.
To do this, we need the following lemma.

\vskip 3mm \noindent
  {\bf Lemma 3.2} \textit{Using the notation above, let $1\leq j\leq r$. Then we have the following conclusions}.

\vskip 2mm\par
  (i) \textit{$\delta_j=1$ for all $j=1,\ldots,\lambda$}.

\vskip 2mm\par
  (ii) \textit{$d_1=1$, and $2\mid d_j$ for all $j=2,\ldots,\lambda$}.

\vskip 3mm\noindent
  \textit{Proof} (i) As $1\leq j\leq \lambda$, we have $\widetilde{f}_j(x)=\delta_jf_j(x)$ where
$\delta_j\in\mathbb{F}_{2^m}^\times$. Since $f_j(x)$ is a monic irreducible divisor of $x^n-1$ in $\mathbb{F}_{2^m}[x]$,
we have
$$f_j(x)=\widetilde{\widetilde{f}_j(x)}=\delta_j \widetilde{f}_j(x)=\delta_j^2f_j(x).$$
This implies $\delta_j^2=1$, and hence $\delta_j=1$ as the characteristic of $\mathbb{F}_{2^m}$ is $2$.

\par
  (ii) Assume that $a\in \mathbb{F}_{2^m}^\times$ and $f(x)=x-a$ is a self-reciprocal polynomial.
Then there exists $\delta\in\mathbb{F}_{2^m}^\times$ such that
$\delta x-\delta a=\delta f(x)=\widetilde{f}(x)=1-ax$. This implies that
$\delta=-a$ and $-\delta a=1$. From this, we deduce that $a^2=1$, and hence $a=1$. Therefore, $f_1(x)=x-1$ is the only
self-reciprocal and monic irreducible divisor of $x^n-1$ in $\mathbb{F}_{2^m}[x]$ with degree $1$.

\par
  Now, let $2\leq j\leq r$. Then $f_j(x)$ is a self-reciprocal and monic irreducible divisor of $x^n-1$ in $\mathbb{F}_{2^m}[x]$ with degree ${\rm deg}(f_j(x))=d_j>1$. This implies that $d_j$ is even from finite field theory.
\hfill $\Box$

\vskip 3mm\par
  Now, all distinct self-dual cyclic
codes over $R$ of length $2n$ can be listed explicitly by Theorem 3.1 and the following theorem.

\vskip 3mm\noindent
  {\bf Theorem 3.3} \textit{Using the notation above, let $1\leq j\leq r$ and $1\leq s\leq k-1$. Then the set
$\Theta_{j,s}$ is determined as follows}.

\begin{description}
\vskip 2mm
\item{(i)}
\textit{If $j=1$, then}
$$\Theta_{1,s}=(\frac{\mathbb{F}_{2^m}[u]}{\langle u^{s}\rangle})^{\times}
=\left\{\sum_{i=0}^{s-1}a_iu^i\mid a_0\neq 0, \ a_i\in \mathbb{F}_{2^m}, i=0,1,\ldots, s-1\right\}$$
\textit{and $|\Theta_{1,s}|=(2^m-1)2^{(s-1)m}$}.

\vskip 2mm
\item{(ii)}
 \textit{Let $2\leq j\leq r$, and $\varrho_j(x)$ is a primitive element of the finite field $\mathcal{F}_j=\frac{\mathbb{F}_{2^m}[x]}{\langle f_j(x)\rangle}$. Then we have}
$$\Theta_{j,1}=\left\{x^{-\frac{d_j}{2}}\varrho_j(x)^{l(2^{\frac{d_j}{2}m}+1)}\mid l=0,1,\ldots,2^{\frac{d_j}{2}m}-2\right\}
\subseteq \mathcal{F}_j;$$
\textit{and for any integer $s$, $2\leq s\leq k-1$, we have}
 $$\Theta_{j,s}=\left\{\sum_{i=0}^{s-1}a_i(x)u^i\mid a_0(x)\in \Theta_{j,1}; \ a_i(x)\in \{0\}\cup \Theta_{j,1},
 1\leq i\leq s-1\right\}.$$
\textit{Therefore, $|\Theta_{j,s}|=(2^{\frac{d_j}{2}m}-1)2^{(s-1)\frac{d_j}{2}m}$
for all $s=1,2,\ldots, k-1$}.
\end{description}

\vskip 3mm\noindent
  \textit{Proof}
   (i) Let $j=1$. By $f_1(x)=x-1$ and Lemma 2.2(ii), we have that $x\equiv 1$ (mod $f_1(x)$),
${\cal F}_1=\frac{\mathbb{F}_{2^m}[x]}{\langle x-1\rangle}=\mathbb{F}_{2^m}$ and
$$({\cal F}_j[u]/\langle u^{s}\rangle)^{\times}=\{\sum_{i=0}^{s-1}a_iu^i\mid a_0\neq 0, \ a_i\in \mathbb{F}_{2^m}, i=0,1,\ldots, s-1\}.$$
In this case, by Lemma 3.2 Condition (3) is simplified to
\begin{equation}
\omega+\widehat{\omega}=\omega+\omega\equiv 0 \ ({\rm mod} \ x-1).
\end{equation}
It is clear that every elements $\omega\in ({\cal F}_j[u]/\langle u^{s}\rangle)^{\times}$ satisfies the above condition. Hence $\Theta_{1,s}=({\cal F}_j[u]/\langle u^{s}\rangle)^{\times}$ and $|\Theta_{1,s}|=(2^m-1)2^{(s-1)m}$.

\vskip 3mm\par
  (ii) Let $2\leq j\leq \lambda$. Then $d_j$ is even and it is well known that
\begin{equation}
x^{-1}=x^{2^{m\frac{d_j}{2}}} \ {\rm in} \ \mathcal{F}_j.
\end{equation}
Let $\omega=\omega(x)\in (\mathcal{F}_j[u]/\langle u^s\rangle)^\times$. By Lemma 2.2(ii), $\omega(x)$
has a unique $u$-expansion:
$$\omega(x)=\sum_{i=0}^{s-1}u^ia_i(x), \ a_0(x)\neq 0, \
a_0(x),a_1(x),\ldots,a_{s-1}(x)\in \mathcal{F}_j.$$
As ${\rm gcd}(x, f_j(x))=1$, by Lemma 3.2(i) the Condition (3)
for $\omega=\omega(x)\in (\mathcal{F}_j[u]/\langle u^s\rangle)^\times$ is transformed to
$x^{\frac{d_j}{2}}\omega(x)+x^{-\frac{d_j}{2}}\omega(x^{-1})\equiv 0 \ ({\rm mod} \ f_j(x))$, i.e,
$$\sum_{i=0}^{s-1}u^i\left(x^{\frac{d_j}{2}}a_i(x)\right)+\sum_{i=0}^{s-1}u^i\left(x^{-\frac{d_j}{2}}a_i(x^{-1})\right)\equiv 0 \ ({\rm mod} \ f_j(x)).$$
This is equivalent to the following congruence equations
\begin{equation}
x^{\frac{d_j}{2}}a_i(x)+x^{-\frac{d_j}{2}}a_i(x^{-1})\equiv 0 \ ({\rm mod} \ f_j(x)),
\ i=0,1,\ldots,s-1.
\end{equation}
For each $0\leq i\leq s-1$, let $\xi_i(x)=x^{\frac{d_j}{2}}a_i(x)\in \mathcal{F}_j$. Then
\begin{equation}
a_i(x)=x^{-\frac{d_j}{2}}\xi_i(x)\in \mathcal{F}_j.
\end{equation}

\par
  For any $b\in \mathcal{F}_j$, by $b^{2^m}=b$ we have $b^{2^{\frac{d_j}{2}m}}=b^{(2^m)^{\frac{d_j}{2}}}=b$ in $\mathcal{F}_j$.
Then by $x^{-1}=x^{2^{\frac{d_j}{2}m}}$ and $\xi_i(x)=x^{\frac{d_j}{2}}a_i(x)$, it follows that
$$x^{-\frac{d_j}{2}}a_i(x^{-1})=\xi_i(x^{-1})=\xi_i(x)^{2^{\frac{d_j}{2}m}}.$$
Therefore, the Condition (6) is equivalent to
$$\xi_i(x)(\xi_i(x)^{2^{\frac{d_j}{2}m}-1}-1)=\xi_i(x)+(\xi_i(x))^{2^{\frac{d_j}{2}m}}=0
\ {\rm in} \ \mathcal{F}_j, \ i=0,1,\ldots, s-1.$$
From the latter condition, we deduce that $\xi_i(x)=0$ when $s\geq 2$ or
$\xi_i(x)\in \mathcal{F}_j$ satisfying $\xi_i(x)^{2^{\frac{d_j}{2}m}-1}=1$ for all $s$.

\par
  Since $\varrho_j(x)$ is a primitive element of $\mathcal{F}_j$, the multiplicative
order of $\varrho_j(x)$ is $2^{d_jm}-1=(2^{\frac{d_j}{2}m}+1)(2^{\frac{d_j}{2}m}-1)$. This implies that
$\varrho_j(x)^{2^{\frac{d_j}{2}m}+1}$ is a primitive $(2^{\frac{d_j}{2}m}-1)$th root of unity. Hence
$\xi_i(x)^{2^{\frac{d_j}{2}m}-1}=1$ if and only if $$\xi_i(x)=(\varrho_j(x)^{2^{\frac{d_j}{2}m}+1})^l=\varrho_j(x)^{l(2^{\frac{d_j}{2}m}+1)}, \ 0\leq l\leq 2^{\frac{d_j}{2}m}-2.$$

\par
  Therefore, the conclusion for $\Omega_{j,s}$ follows from Equation (7) immediately.
Moreover, we have $|\Theta_{j,s}|=|\Theta_{j,1}|\prod_{i=2}^{s}(|\Theta_{j,1}|+1)=(2^{\frac{d_j}{2}m}-1)2^{(s-1)\frac{d_j}{2}m}$
for all $s=1,2,\ldots, k-1$.
\hfill $\Box$

\vskip 3mm \par
  Now is the time to give an explicit formula
to count the number of all
distinct self-dual cyclic
codes over the ring $R$ of length $2n$.

\vskip 3mm \noindent
  {\bf Corollary 3.4} \textit{Let $\mathcal{N}_S(2^m,k,n)$ be the number of all
distinct self-dual cyclic
codes over the ring $R$ of length $2n$. Then}
$$\mathcal{N}_S(2^m,k,n)=(\sum_{s=0}^{\frac{k}{2}}2^{ms})(\prod_{j=2}^{\lambda}\sum_{s=0}^{\frac{k}{2}}2^{\frac{d_j}{2}ms})
(\prod_{j=\lambda+1}^{\lambda+\epsilon}N_{(2^m,d_j,k)}),
\ {\rm when} \ 2\mid k;$$
$$\mathcal{N}_S(2^m,k,n)=(\sum_{s=0}^{\frac{k-1}{2}}2^{ms})(\prod_{j=2}^{\lambda}\sum_{s=0}^{\frac{k-1}{2}}2^{\frac{d_j}{2}ms})
(\prod_{j=\lambda+1}^{\lambda+\epsilon}N_{(2^m,d_j,k)}),
\ {\rm when} \ 2\nmid k.$$

\vskip 3mm \noindent
 \textit{Proof} Let $k$ be even and $1\leq j\leq \lambda$. Then
the number of ideals $C_j$ listed in ($\dag$) of Theorem 3.1(i) is equal to
\begin{eqnarray*}
N_j&=&2+\sum_{t=0}^{\frac{k}{2}-1}|\Theta_{1,\frac{k}{2}-t}|+\sum_{i=\frac{k}{2}+1}^{k-1}|\Theta_{1,k-i}|+k-1-\frac{k}{2}
+\sum_{i=\frac{k}{2}+1}^{k-1}\sum_{t=1}^{k-1-i}|\Theta_{1,k-i-t}|\\
&=&1+\frac{k}{2}+\frac{k}{2}|\Theta_{1,1}|+(\frac{k}{2}-1)|\Theta_{1,2}|+(\frac{k}{2}-2)|\Theta_{1,3}|+2|\Theta_{1,\frac{k}{2}-1}|
+|\Theta_{1,\frac{k}{2}}|\\
&=&1+\frac{k}{2}+\sum_{s=1}^{\frac{k}{2}}(\frac{k}{2}-s+1)|\Theta_{j,s}|.
\end{eqnarray*}

\par
   When $j=1$, we have $|\Theta_{1,s}|=(2^m-1)2^{(s-1)m}=2^{sm}-2^{(s-1)m}$ by Theorem 3.3(i).
From this we deduce that
$$N_1=1+\frac{k}{2}+\sum_{s=1}^{\frac{k}{2}}(\frac{k}{2}-s+1)(2^{sm}-2^{(s-1)m})=\sum_{s=0}^{\frac{k}{2}}2^{ms}.$$

\par
  When $2\leq j\leq \lambda$, we have $|\Theta_{j,s}|=(2^{\frac{d_j}{2}m}-1)2^{(s-1)\frac{d_j}{2}m}=2^{s\frac{d_j}{2}m}-2^{(s-1)\frac{d_j}{2}m}$ by Theorem 3.3(ii).
From this we deduce that
$$N_j=1+\frac{k}{2}+\sum_{s=1}^{\frac{k}{2}}(\frac{k}{2}-s+1)(2^{s\frac{d_j}{2}m}-2^{(s-1)\frac{d_j}{2}m})
=\sum_{s=0}^{\frac{k}{2}}2^{\frac{d_j}{2}ms}.$$

\par
  Then by Theorem 3.1 we conclude that
$$\mathcal{N}_S(2^m,k,n)=N_1(\prod_{j=2}^\lambda N_j)(\prod_{j=\lambda+1}^{\lambda+\epsilon}N_{(2^m,d_j,k)}).$$

\par
  The conclusion for odd integer $k\geq 3$ can by proved similarly. Here, we omit it.
\hfill $\Box$


\section{Self-dual and $2$-quasi-cyclic codes of length $4n$ over $\mathbb{F}_{2^m}$ derived from self-deal cyclic codes
of length $2n$ over $\mathbb{F}_{2^m}+u\mathbb{F}_{2^m}$}
\noindent
 In this section, We focus on investigating self-dual cyclic codes of length $2n$ over
$R=\mathbb{F}_{2^m}+u\mathbb{F}_{2^m}$ $(u^2=0)$, where $n$ is odd.
By Lemma 2.3, Corollary 2.5, Theorem 3.1 and Theorem 3.3, we obtain the following conclusion.

\vskip 3mm\noindent
  {\bf Corollary 4.1} \textit{The number of self-dual cyclic codes of length $2n$
over the ring $\mathbb{F}_{2^m}+u\mathbb{F}_{2^m}$ $(u^2=0)$ is}
$$(1+2^m)\cdot\prod_{j=2}^\lambda(1+2^{\frac{d_j}{2}m})\cdot\prod_{j=\lambda+1}^{\lambda+\epsilon}(5+2^{d_jm}).$$
\textit{Precisely, all these codes are given by}
$${\cal C}=\left(\oplus_{j=1}^\lambda \varepsilon_j(x)C_j\right)\oplus
\left(\oplus_{j=\lambda+1}^{\lambda+\epsilon}(\varepsilon_{j}(x)C_{j}\oplus\varepsilon_{j+\epsilon}(x)C_{j+\epsilon})\right),$$
\textit{where $C_j$ is an ideal of ${\cal K}_j+u{\cal K}_j$ $(u^2=0)$ listed as follows}:

\begin{description}
\vskip 2mm
\item{(i)}
 \textit{$C_1$ is one of the following $1+2^m$ ideals}:
\par
  \textit{$\langle u\rangle$, $\langle (x-1)\rangle$,  $\langle u+(x-1)\omega\rangle$ where $\omega\in \mathbb{F}_{2^m}$ and $\omega\neq 0$}.

\vskip 2mm
\item{(ii)}
\textit{Let $2\leq j\leq \lambda$, and $\varrho_j(x)$ is a primitive element of the finite field $\mathcal{F}_j=\frac{\mathbb{F}_{2^m}[x]}{\langle f_j(x)\rangle}$. Then $C_j$ is one of the following $1+2^{\frac{d_j}{2}m}$ ideals}:

\par
  \textit{$\langle u\rangle$, $\langle f_j(x)\rangle$};

\par
   \textit{$\langle u+f_j(x)\omega(x)\rangle$, where $\omega(x)\in \Theta_{j,1}$ and}
   $$\Theta_{j,1}=\left\{x^{-\frac{d_j}{2}}\varrho_j(x)^{l(2^{\frac{d_j}{2}m}+1)}
   \ ({\rm mod} \ f_j(x)) \ \mid l=0,1,\ldots,2^{\frac{d_j}{2}m}-2\right\}.$$

\item{(iii)}
 \textit{Let $\lambda+1\leq j\leq \lambda+\epsilon$. Then the pair $(C_j,C_{j+\epsilon})$ of ideals is one of the following $5+2^{d_jm}$ cases in the following table}:
\begin{center}
\begin{tabular}{llll}\hline
$\mathcal{L}$ &  $C_j$ (mod $f_j(x)^2$) & $|C_j|$ & $C_{j+\epsilon}$  (mod $f_{j+\epsilon}(x)^2$)\\ \hline
$3$ & $\bullet$ $\langle u^i\rangle$ \ $(i=0,1,2)$  & $4^{(2-i)d_jm}$ & $\diamond$ $\langle u^{2-i}\rangle$ \\
$2$ & $\bullet$  $\langle u^sf_j(x)\rangle$ ($s=0,1$) & $2^{(2-s)d_jm}$ & $\diamond$ $\langle u^{2-s},f_{j+\epsilon}(x)\rangle$ \\
$2^{d_jm}-1$ & $\bullet$ $\langle u+f_j(x)\omega\rangle$ & $4^{d_jm}$ & $\diamond$ $\langle u+f_{j+\epsilon}(x)\omega^{\prime}\rangle$ \\
$1$ & $\bullet$ $\langle u,f_j(x)\rangle$ & $2^{3d_jm}$ & $\diamond$ $\langle uf_{j+\epsilon}(x)\rangle$ \\
 \hline
\end{tabular}
\end{center}
\textit{where}

\begin{description}
\item{}
 \textit{$\mathcal{L}$ is the number of pairs $(C_j,C_{j+\epsilon})$ in the same row};

\item{}
  \textit{$\omega=\omega(x)\in \mathcal{F}_j
  =\{\sum_{i=0}^{d_j-1}a_ix^i\mid a_0,a_1,\ldots,a_{d_j-1}\in \mathbb{F}_{2^m}\}$ and $\omega\neq 0$},

\item{}
 $\omega^{\prime}=\delta_jx^{-d_j}\omega(x^{-1})$ $({\rm mod} \ f_{j+\epsilon}(x))$.
\end{description}
\end{description}

\vskip 3mm\par
  For the cases of
$k=3,4,5$, we list all
distinct self-dual cyclic codes  of
length $2n$ over the ring $\frac{\mathbb{F}_{2^m}[u]}{\langle u^k\rangle}$ in Appendix of this paper.

\vskip 3mm\par
   Now, let's consider how to calculate the number of self-dual cyclic codes
 of length $2n$ over the ring
$R=\mathbb{F}_{2^m}+u \mathbb{F}_{2^m}$ directly from the odd positive integer $n$. Let $J_1,J_2,\ldots,J_r$ be all the distinct
$2^m$-cyclotomic cosets modulo $n$ corresponding to the factorization $x^n-1=f_1(x)f_2(x)
\ldots f_r(x)$, where $f_1(x)=x-1, f_2(x),\ldots,  f_r(x)$ are distinct monic
irreducible polynomials in $\mathbb{F}_{2^m}[x]$. Then we have $r=\lambda+2\epsilon$ and

\begin{enumerate}
  \item[$\bullet$] $J_1=\{0\}$, the set $J_j$ satisfies $J_j=-J_j \pmod{n}$ and $|J_j|=d_j$ for all $j=2,\ldots,\lambda$;

  \item[$\bullet$] $J_{\lambda+l+\epsilon}=-J_{\lambda+l} \pmod{n}$ and $|J_{\lambda+l}|=|J_{\lambda+l+\epsilon}|=d_{\lambda+l}$, for all $l=1,\ldots,\epsilon$.
  \end{enumerate}

\noindent
From this and by Corollary 4.1, we deduce that
the number of self-dual cyclic codes over $R$ of length
$2n$ is
$$(1+2^m)\cdot\prod_{j=2}^\lambda(1+2^{\frac{|J_i|}{2}m})\cdot\prod_{j=\lambda+1}^{\lambda+\epsilon}(5+2^{|J_i|m}).$$

\par
  As an example, we list the number $\mathcal{N}$ of self-dual cyclic codes over $\mathbb{F}_2+u\mathbb{F}_2$ of length $2n$, where
$n$ is odd and $6 \leq 2n \leq 98$, in the following table.
\begin{center}
{\small
\begin{tabular}{ll|ll}\hline
  $2n$ & $\mathcal{N}$  &   $2n$ & $\mathcal{N}$ \\ \hline
 $6$ & $9=3(1+2)$   &  $54$ & $41553=3(1+2)(1+2^3)(1+2^9)$ \\
 $10$ & $15=3(1+2^2)$  & $58$ & $49155=3(1+2^{14})$ \\
 $14$ & $39=3(5+2^3)$ & $62$ & $151959=3(5+2^5)^3$ \\
 $18$ & $81=3(1+2)(1+2^3)$ & $66$ & $323433=3(1+2)(1+2^5)^3$ \\
 $22$ & $99=3(1+2^5)$ & $70$ & $799695=3(1+2^2)(5+2^3)(5+2^{12})$ \\
 $26$ & $195=3(1+2^6)$ & $74$ & $786435=3(1+2^{18})$ \\
 $30$ & $945=3(1+2)(1+2^2)(5+2^4)$ & $78$ & $2399085=3(1+2)(1+2^6)(5+2^{12})$\\
 $34$ & $867=3(1+2^4)^2$            & $82$     & $3151875=3(1+2^{10})^2$  \\
 $38$ & $1539=3(1+2^9)$                & $86$     & $6440067=3(1+2^7)^3$  \\
 $42$ & $8073=3(1+2)(5+2^3)(5+2^6)$ & $90$     & $34879005$  \\
 $46$ & $6159=3(5+2^{11})$            & $94$     & $25165839=3(5+2^{23})$  \\
 $50$ & $15375=3(1+2^2)(1+2^{10})$   & $98$     & $81789123=3(5+2^3)(5+2^{21})$  \\
\hline
\end{tabular} }
\end{center}
where $34879005=3(1+2)(1+2^2)(1+2^3)(5+2^4)(5+2^{12})$.

\vskip 3mm \par
  Then we consider how to construct
self-dual and $2$-quasi-cyclic codes of length $4n$ over $\mathbb{F}_{2^m}$ from
self-dual cyclic codes of length $2n$ over $\mathbb{F}_{2^m}+u\mathbb{F}_{2^m}$.

\par
  Let $\alpha=a+bu\in R$ where $a,b\in \mathbb{F}_{2^m}$. As in [4], we define
$\phi(\alpha)=(b,a+b)$ and define the \textit{Lee weight} of $\alpha$ by
${\rm w}_L(\alpha)={\rm w}_H(b,a+b)$, where ${\rm w}_H(b,a+b)$ is the Hamming weight of the vector
$(b,a+b)\in \mathbb{F}_{2^m}^2$. Then $\phi$ is an isomorphism of $\mathbb{F}_{2^m}$-linear
spaces from $R$ onto $\mathbb{F}_{2^m}^2$, and can be extended to an isomorphism of $\mathbb{F}_{2^m}$-linear
spaces from $\frac{R[x]}{\langle x^{2n}-1\rangle}$ onto $\mathbb{F}_{2^m}^{4n}$ by the rule:
\begin{equation}
\phi(\xi)=(b_0,b_1,\ldots,b_{2n-1},a_0+b_0,a_1+b_1,\ldots,a_{2n-1}+b_{2n-1}),
\end{equation}
for all $\xi=\sum_{i=0}^{2n-1}\alpha_ix^i\in \frac{R[x]}{\langle x^{2n}-1\rangle}$, where $\alpha_i=a_i+b_iu$ with $a_i,b_i\in \mathbb{F}_{2^m}$ and $i=0,1,\ldots,2n-1$.

\par
  The following conclusion can be derived from Corollary 14 of [4].

\vskip 3mm\noindent
  {\bf Lemma 4.2} \textit{Using the notation above, let $\mathcal{C}$ be an ideal of the ring $\frac{R[x]}{\langle x^{2n}-1\rangle}$ and set $\phi(\mathcal{C})=\{\phi(\xi)\mid \xi\in \mathcal{C}\}\subseteq \mathbb{F}_{2^m}^{4n}$. Then}

\begin{description}
\item{(i)}
  \textit{$\phi(\mathcal{C})$ is a $2$-quasi-cyclic code over $\mathbb{F}_{2^m}$ of length $4n$, i.e.,}
$$(b_{2n-1},b_0,b_1,\ldots,b_{2n-2},c_{2n-1},c_0,c_1,\ldots,c_{2n-2})\in \phi(\mathcal{C})$$
\textit{for all $(b_0,b_1,\ldots,b_{2n-2},b_{2n-1},c_0,c_1,\ldots,c_{2n-2},c_{2n-1})\in \phi(\mathcal{C})$}.

\vskip 2mm
\item{(ii)}
 \textit{The Hamming weight distribution of $\phi(\mathcal{C})$ is exactly the same as the Lee weight distribution of $\mathcal{C}$}.

\vskip 2mm
\item{(iii)}
  \textit{$\phi(\mathcal{C})$ is a self-dual code over $\mathbb{F}_{2^m}$ of length $4n$ if
$\mathcal{C}$ is a self-dual code over $R$ of length $2n$}.
\end{description}

\par
  By Corollary 4.1, we can get a class of self-dual and $2$-quasi-cyclic codes over $\mathbb{F}_{2^m}$ of length $4n$
from the class of self-dual cyclic code over $R$ of length $2n$ and the Gray map $\phi$ defined by Equation (8).
In the following, we consider how to give an efficient encoder for each
self-dual and $2$-quasi-cyclic code $\phi(\mathcal{C})$ of length $4n$ over $\mathbb{F}_{2^m}$ derived
from a self-dual cyclic code $\mathcal{C}$ of length $2n$ over $\mathbb{F}_{2^m}+u\mathbb{F}_{2^m}$.

\par
  To simplify the symbol, in the following we identify each polynomial
$a(x)=a_0+a_1x+\ldots+a_{2n-1}x^{2n-1}\in\frac{\mathbb{F}_{2^m}[x]}{\langle x^{2n}-1\rangle}$
with the vector $(a_0,a_1,\ldots,a_{2n-1})\in \mathbb{F}_{2^m}^{2n}$. Moreover, for any integer $1\leq s\leq n-1$ we
denote:
\begin{equation}
[a(x)]_s=\left(\begin{array}{c}a(x)\cr xa(x)\cr \ldots \cr x^{s-1}a(x)\end{array}\right)
=\left(\begin{array}{ccccc}a_0 & a_1 &\ldots & a_{2n-2} & a_{2n-1}\cr
a_{2n-1} & a_0 &\ldots & a_{2n-3} & a_{2n-2}\cr
\ldots &\ldots &\ldots &\ldots &\ldots \cr
a_{2n-s+1} & a_{2n-s+2} & \ldots & a_{2n-s-1} & a_{2n-s}\end{array}\right)
\end{equation}
which is a matrix over $\mathbb{F}_{2^m}$ of size $s\times 2n$.

\vskip 3mm \noindent
  {\bf Theorem 4.3} \textit{Using the notation above,
each self-dual and $2$-quasi-cyclic code $\phi(\mathcal{C})$ of length $4n$ over $\mathbb{F}_{2^m}$ derived
from a self-dual cyclic code $\mathcal{C}$ of length $2n$ over $\mathbb{F}_{2^m}+u\mathbb{F}_{2^m}$
has a $\mathbb{F}_{2^m}$-generator matrix given by:
$$G=\left(\begin{array}{c}G_1 \cr G_2 \cr \ldots \cr G_{\lambda+\epsilon}\end{array}\right)$$
in which for each integer $j$, $1\leq j\leq r$, $G_j$ is a
matrix over $\mathbb{F}_{2^m}$ listed in the following}:

\begin{description}
\vskip 2mm
\item{(i)}
  \textit{$G_1$ is one of the following $1+2^m$ matrices with size $2\times 4n$}:

\vskip 2mm\par
  $\left(\begin{array}{cc} \varepsilon_1(x) & \varepsilon_j(x)\cr (x-1)\varepsilon_1(x) & (x-1)\varepsilon_1(x)\end{array}\right)$,
$\left(\begin{array}{cc} 0 & (x-1)\varepsilon_1(x) \cr (x-1)\varepsilon_1(x) & (x-1)\varepsilon_1(x)\end{array}\right)$,

\par
  $\left(\begin{array}{cc} \varepsilon_1(x) & \varepsilon_1(x)+(x-1)\varepsilon_1(x)\omega\cr (x-1)\varepsilon_1(x) & (x-1)\varepsilon_1(x)\end{array}\right)$ \textit{where $\omega\in \mathbb{F}_{2^m}$ and $\omega\neq 0$}.

\vskip 2mm
\item{(ii)}
  \textit{Let $2\leq j\leq \lambda$. Then $G_j$ is one of the following $1+2^{\frac{d_j}{2}m}$ matrices
with size $2d_j\times 4n$}:

\vskip 2mm\par
  $\left(\begin{array}{cc} [\varepsilon_j(x)]_{d_j} & [\varepsilon_j(x)]_{d_j}\cr [f_j(x)\varepsilon_j(x)]_{d_j} & [f_j(x)\varepsilon_j(x)]_{d_j} \end{array}\right)$,
$\left(\begin{array}{cc} 0 & [f_j(x)\varepsilon_j(x)]_{d_j} \cr [f_j(x)\varepsilon_j(x)]_{d_j} & [f_j(x)\varepsilon_j(x)]_{d_j}\end{array}\right)$,

\par
  $\left(\begin{array}{cc} [\varepsilon_j(x)]_{d_j} & [(1+f_j(x)\omega(x))\varepsilon_j(x)]_{d_j}
  \cr [f_j(x)\varepsilon_j(x)]_{d_j} & [f_j(x)\varepsilon_j(x)]_{d_j}\end{array}\right)$ \textit{where} $$\omega(x)=x^{-\frac{d_j}{2}}\varrho_j(x)^{l(2^{\frac{d_j}{2}m}+1)} \ ({\rm mod} \ f_j(x)),
  \ l=0,1,\ldots,2^{\frac{d_j}{2}m}-2.$$

\vskip 2mm
\item{(iii)}
   \textit{Let $\lambda+1\leq j\leq \lambda+\epsilon$. Then $G_j$ is one of the following $5+2^{d_jm}$ matrices
with size $4d_j\times 4n$}:

\vskip 2mm\noindent
 {\small   $\left(\begin{array}{cc} 0 & [\varepsilon_j(x)]_{d_j} \cr [\varepsilon_j(x)]_{d_j} & [\varepsilon_j(x)]_{d_j} \cr
  0 & [f_j(x)\varepsilon_j(x)]_{d_j} \cr [f_j(x)\varepsilon_j(x)]_{d_j} & [f_j(x)\varepsilon_j(x)]_{d_j}\end{array}\right)$,
 $\left(\begin{array}{cc} [\varepsilon_j(x)]_{d_j} & [\varepsilon_j(x)]_{d_j}\cr [f_j(x)\varepsilon_j(x)]_{d_j} & [f_j(x)\varepsilon_j(x)]_{d_j} \cr
 [\varepsilon_{j+\epsilon}(x)]_{d_j} & [\varepsilon_{j+\epsilon}(x)]_{d_j}\cr [f_{j+\epsilon}(x)\varepsilon_{j+\epsilon}(x)]_{d_j} & [f_{j+\epsilon}(x)\varepsilon_{j+\epsilon}(x)]_{d_j}\end{array}\right)$;  }

\noindent
  {\small   $\left(\begin{array}{cc} 0 & [\varepsilon_{j+\epsilon}(x)]_{d_j} \cr [\varepsilon_{j+\epsilon}(x)]_{d_j} & [\varepsilon_{j+\epsilon}(x)]_{d_j} \cr
  0 & [f_{j+\epsilon}(x)\varepsilon_{j+\epsilon}(x)]_{d_j} \cr [f_{j+\epsilon}(x)\varepsilon_{j+\epsilon}(x)]_{d_j} & [f_{j+\epsilon}(x)\varepsilon_{j+\epsilon}(x)]_{d_j}\end{array}\right)$;}

\noindent
  {\small   $\left(\begin{array}{cc} 0 & [f_j(x)\varepsilon_j(x)]_{d_j} \cr [f_j(x)\varepsilon_j(x)]_{d_j} & [f_j(x)\varepsilon_j(x)]_{d_j}\cr  0 & [f_{j+\epsilon}(x)\varepsilon_{j+\epsilon}(x)]_{d_j}
  \cr [f_{j+\epsilon}(x)\varepsilon_{j+\epsilon}(x)]_{d_j} & [f_{j+\epsilon}(x)\varepsilon_{j+\epsilon}(x)]_{d_j}\end{array}\right)$,

   $\left(\begin{array}{cc} [f_j(x)\varepsilon_j(x)]_{d_j} & [f_j(x)\varepsilon_j(x)]_{d_j} \cr
   [\varepsilon_{j+\epsilon}(x)]_{d_j} & [\varepsilon_{j+\epsilon}(x)]_{d_j}\cr [f_{j+\epsilon}(x)\varepsilon_{j+\epsilon}(x)]_{d_j} & [f_{j+\epsilon}(x)\varepsilon_{j+\epsilon}(x)]_{d_j}
   \cr  0 & [f_{j+\epsilon}(x)\varepsilon_{j+\epsilon}(x)]_{d_j} \end{array}\right)$;  }

\noindent
  {\small $\left(\begin{array}{cc} [\varepsilon_j(x)]_{d_j} & [(1+f_j(x)\omega(x))\varepsilon_j(x)]_{d_j}
  \cr [f_j(x)\varepsilon_j(x)]_{d_j} & [f_j(x)\varepsilon_j(x)]_{d_j} \cr
  [\varepsilon_{j+\epsilon}(x)]_{d_j} & [(1+f_{j+\epsilon}(x)\omega^\prime(x))\varepsilon_{j+\epsilon}(x)]_{d_j}
  \cr [f_{j+\epsilon}(x)\varepsilon_{j+\epsilon}(x)]_{d_j} & [f_{j+\epsilon}(x)\varepsilon_{j+\epsilon}(x)]_{d_j}\end{array}\right)$

\textit{where $\omega^\prime(x)=\delta_jx^{-d_j}w(x^{-1})$ $({\rm mod} \ f_{j+\epsilon}(x))$,
$\omega(x)\in \mathcal{F}_j$ and $\omega(x)\neq 0$};   }

\noindent
  {\small $\left(\begin{array}{cc}
   [\varepsilon_j(x)]_{d_j} & [\varepsilon_j(x)]_{d_j}\cr [f_j(x)\varepsilon_j(x)]_{d_j} & [f_j(x)\varepsilon_j(x)]_{d_j}
   \cr  0 & [f_j(x)\varepsilon_j(x)]_{d_j}
   \cr [f_{j+\epsilon}(x)\varepsilon_{j+\epsilon}(x)]_{d_j} & [f_{j+\epsilon}(x)\varepsilon_{j+\epsilon}(x)]_{d_j}
   \end{array}\right)$.}
\end{description}

\vskip 3mm\noindent
  \textit{Proof} Let $\mathcal{C}$ be a self-dual cyclic codes
over $\mathbb{F}_{2^m}+u\mathbb{F}_{2^m}$ of length $2n$. By Corollary 4.1, $\mathcal{C}$ has
a unique direct decomposition:
\begin{equation}
\mathcal{C}=\mathcal{C}_1\oplus \mathcal{C}_2\oplus\ldots\oplus \mathcal{C}_{\lambda+\epsilon},
\end{equation}
\textit{where $\mathcal{C}_j=\varepsilon_j(x)C_j=\{\varepsilon_j(x)\xi\mid \xi\in C_j\}$ $({\rm mod} \ x^{2n}-1)$
for all $j=1,\ldots,\lambda$, $\mathcal{C}_j=\varepsilon_j(x)C_j\oplus \varepsilon_{j+\epsilon}(x)C_{j+\epsilon}$ $({\rm mod} \ x^{2n}-1)$
for all $j=\lambda+1,\ldots, \lambda+\epsilon$, and}

\begin{itemize}
\item
  $C_1$ is given by
Corollary 4.1(i);

\item
  $C_j$ is given by Corollary 4.1(ii) for all $j=2,\ldots \lambda$;

\item
  $(C_j, C_{j+\epsilon})$ is given by
Corollary 4.1(iii) for all $j=\lambda+1,\ldots, \lambda+\epsilon$.
\end{itemize}

\par
  Now, let $\alpha(x)$ be an arbitrary element in the ring $\mathcal{K}_j+u\mathcal{K}_j$ ($u^2=0$) where
$\mathcal{K}_j=\frac{\mathbb{F}_{2^m}[x]}{\langle f_j(x)^2\rangle}$. Then there is a unique
tuple $(\alpha_0,\alpha_1,\alpha_2,\alpha_3)$ of elements in $\mathcal{F}_j=\frac{\mathbb{F}_{2^m}[x]}{\langle f_j(x)\rangle}$
such that
\begin{equation}
\alpha=\left(\alpha_0+\alpha_1f_j(x)\right)+u\left(\alpha_2+\alpha_3f_j(x)\right).
\end{equation}
Since $\{1,x,\ldots,x^{d_j-1}\}$ is an $\mathbb{F}_{2^m}$-basis of $\mathcal{F}_j$, for each integer
$0\leq t\leq 3$ there is a unique row matrix $\underline{a}_t=(a_{t,0},a_{t,1},\ldots,a_{t,d_j-1})\in \mathbb{F}_{2^m}^{d_j}$
such that
\begin{equation}
\alpha_t=a_{t,0}+a_{t,1}x+\ldots+a_{t,d_j-1}x^{d_j-1}=\underline{a}_t X_{d_j}, \
{\rm where} \ X_{d_j}=\left(\begin{array}{c}1 \cr x \cr \ldots \cr x^{d_j-1}\end{array}\right).
\end{equation}

\par
  (ii) Let $2\leq j\leq \lambda$ and $C_j$ be an ideal of the ring $\mathcal{K}_j+u\mathcal{K}_j$ given by Corollary 4.1(ii). Then we have one of the following three cases.

\par
  (ii-1) $C_j=\langle u\rangle$. In this case, by Equation (11) and $u^2=0$ we see that every
element of $C_j$ is given by: $\xi = \alpha u=\left(\alpha_0+\alpha_1f_j(x)\right)u$, where
$\alpha_0,\alpha_1\in \mathcal{F}_j$. Then by Equations (12) and (9) we have
\begin{eqnarray*}
\phi(\varepsilon_j(x)\xi) &=&(\alpha_0\varepsilon_j(x)+\alpha_1f_j(x)\varepsilon_j(x),
\alpha_0\varepsilon_j(x)+\alpha_1f_j(x)\varepsilon_j(x))\\
 &=&(\underline{a}_0 X_{d_j}\varepsilon_j(x)+\underline{a}_1 X_{d_j}f_j(x)\varepsilon_j(x),
\underline{a}_0 X_{d_j}\varepsilon_j(x)+\underline{a}_1 X_{d_j}f_j(x)\varepsilon_j(x))\\
 &=& (\underline{a}_0,\underline{a}_1)G_j,
\end{eqnarray*}
where $G_j=\left(\begin{array}{cc} [\varepsilon_j(x)]_{d_j} & [\varepsilon_j(x)]_{d_j} \cr
 [f_j(x)\varepsilon_j(x)]_{d_j} & [f_j(x)\varepsilon_j(x)]_{d_j}\end{array}\right)\in {\rm M}_{2d_j\times 4n}$.
On the other hand, we know that ${\rm dim}_{\mathbb{F}_{2^m}}(\phi(\mathcal{C}_j))={\rm dim}_{\mathbb{F}_{2^m}}(C_j)=2d_j$
by Lemma 2.3.
From these, we deduce that $G_j$ is a generator matrix of the $\mathbb{F}_{2^m}$-linear
code $\mathcal{C}_j$.

\par
  (ii-2) $C_j=\langle f_j(x)\rangle$. In this case, by Equation (11) and $f_j(x)^2=0$ we see that every
element of $C_j$ is given by: $\xi = \alpha f_j(x)=\alpha_0f_j(x)+u\alpha_2f_j(x)$, where
$\alpha_0,\alpha_2\in \mathcal{F}_j$. Then by Equations (12) and (9) we have
\begin{eqnarray*}
\phi(\varepsilon_j(x)\xi) &=&(\alpha_2f_j(x)\varepsilon_j(x),
\alpha_0f_j(x)\varepsilon_j(x)+\alpha_2f_j(x)\varepsilon_j(x))\\
 &=&(\underline{a}_2 X_{d_j}f_j(x)\varepsilon_j(x),
\underline{a}_0 X_{d_j}f_j(x)\varepsilon_j(x)+\underline{a}_2 X_{d_j}f_j(x)\varepsilon_j(x))\\
 &=& (\underline{a}_0,\underline{a}_2)G_j,
\end{eqnarray*}
where $G_j=\left(\begin{array}{cc} 0 & [f_j(x)\varepsilon_j(x)]_{d_j} \cr
 [f_j(x)\varepsilon_j(x)]_{d_j} & [f_j(x)\varepsilon_j(x)]_{d_j}\end{array}\right)\in {\rm M}_{2d_j\times 4n}$.
On the other hand, we know that ${\rm dim}_{\mathbb{F}_{2^m}}(\phi(\mathcal{C}_j))={\rm dim}_{\mathbb{F}_{2^m}}(C_j)=2d_j$
by Lemma 2.3.
From these, we deduce that $G_j$ is a generator matrix of the $\mathbb{F}_{2^m}$-linear
code $\mathcal{C}_j$.

\par
  (ii-3) $C_j=\langle u+f_j(x)\omega(x)\rangle$ where $\omega(x)=x^{-\frac{d_j}{2}}\varrho_j(x)^{l(2^{\frac{d_j}{2}m}+1)}
  \in \mathcal{F}_j^\times$
and $0\leq l\leq 2^{\frac{d_j}{2}m}-2$. In this case, by Equation (11) and $u^2=f_j(x)^2=0$ we see that every
element of $C_j$ is given by:
$$\xi = \alpha \left(u+f_j(x)\omega(x)\right)=\alpha_0f_j(x)\omega(x)
+u\left(\alpha_0+\alpha_1f_j(x)+\alpha_2f_j(x)\omega(x)\right),$$ where
$\alpha_0,\alpha_1,\alpha_2\in \mathcal{F}_j$. Denote $\beta=\alpha_1+\omega(x)\alpha_2$.
Then $\beta\in \mathcal{F}_j$ and can be written as $\beta=\underline{b}X_{d_j}$ where
$\underline{b}\in \mathbb{F}_{2^m}^{d_j}$.
Now, by Equations (12) and (9) we have
\begin{eqnarray*}
\phi(\varepsilon_j(x)\xi) &=&\varepsilon_j(x)(\alpha_0+\alpha_1f_j(x)+\alpha_2f_j(x)\omega(x),\\
 && \alpha_0f_j(x)\varepsilon_j(x)\omega(x)+\alpha_0+\alpha_1f_j(x)+\alpha_2f_j(x)\omega(x))\\
 &=&\varepsilon_j(x)(\alpha_0+\beta f_j(x), \alpha_0(1+f_j(x)\varepsilon_j(x)\omega(x))+\beta f_j(x))\\
  &=&(\alpha_0,\beta)
    \left(\begin{array}{cc} \varepsilon_j(x) & (1+f_j(x)\omega(x))\varepsilon_j(x) \cr f_j(x)\varepsilon_j(x) & f_j(x)\varepsilon_j(x)
   \end{array}\right)\\
 &=& (\underline{a}_0,\underline{b})G_j,
\end{eqnarray*}
where $G_j=\left(\begin{array}{cc} [\varepsilon_j(x)]_{d_j} & [(1+f_j(x)\omega(x))\varepsilon_j(x)]_{d_j} \cr
 [f_j(x)\varepsilon_j(x)]_{d_j} & [f_j(x)\varepsilon_j(x)]_{d_j}\end{array}\right)\in {\rm M}_{2d_j\times 4n}$.
On the other hand, we have ${\rm dim}_{\mathbb{F}_{2^m}}(\phi(\mathcal{C}_j))={\rm dim}_{\mathbb{F}_{2^m}}(C_j)=2d_j$
by Lemma 2.3.
From these, we deduce that $G_j$ is a generator matrix of the $\mathbb{F}_{2^m}$-linear
code $\mathcal{C}_j$.

\par
  The conclusions in (i) and (iii) can be proved similarly as that in (ii) above. Here, we omit these
details.
\hfill $\Box$

\vskip 3mm\par
   As the end of this section, we list all distinct self-dual cyclic codes $\mathcal{C}$ over $\mathbb{F}_2+u\mathbb{F}_2$ of length $30$. We have
   $$x^{15}-1=f_1(x)f_2(x)f_3(x)f_4(x)f_5(x),$$
where

\begin{itemize}
\item
 $f_1(x)=x-1$, $f_2(x)=x^2+x+1$, $f_3(x)=x^4+x^3+x^2+x=1$, \\

\item
 $f_4(x)=x^4+x+1$ and $f_5(x)=x^4+x^3+1$
\end{itemize}
 are irreducible polynomials in $\mathbb{F}_2[x]$
satisfying $\widetilde{f}_j(x)=f_j(x)$ for all $j=1,2,3$, and $\widetilde{f}_4(x)=f_5(x)$ with $\delta_4=1$.
Hence $r=5$, $\lambda=3$, $\epsilon=1$, $d_1=1$, $d_2=2$ and $d_3=d_4=d_5=4$.

\par
  Using the notation in Section 2, we have
\[  \begin{array}{rl}
  \varepsilon_1(x)= & {x}^{28}+{x}^{26}+{x}^{24}+{x}^{22}+{x}^{20}+{x}^{18}
+{x}^{16}+{x}^{14}+{x}^{12}+{x}^{10}+{x}^{8}+{x}^{6}\\
 & +{x}^{4}+x^{2}+1, \\

  \varepsilon_2(x)= & {x}^{28}+{x}^{26}+{x}^{22}+{x}^{20}+{x}^{16}+{x}
^{14}+{x}^{10}+{x}^{8}+{x}^{4}+x^{2}, \\

  \varepsilon_3(x)= & {x}^{28}+{x}^{26}+{x}^{24}+{x}^{22}+{x}^{18}+{x}^{16}+{x}^{14}+{x}^{12}+{x
}^{8}+{x}^{6}+{x}^{4}+x^{2}, \\

 \varepsilon_4(x)= & {x}^{24}+{x}^{18}+{x}^{16}+{x}^{12}+{x}^{8}+
{x}^{6}+{x}^{4}+x^{2}, \\

  \varepsilon_5(x)= & {x}^{28}+{x}^{26}+{x}^{24}+{x}^{22}+{x}^{18}+{x}^{14}+{x}^{12}+{x}^{6}.
\end{array}
\]

\par
  Obviously, $x$ is a primitive element of the finite field $\mathcal{F}_2=\frac{\mathbb{F}_2[x]}{\langle f_2(x)\rangle}$ and
$$\Omega_{2,1}=\{x^{-\frac{2}{2}}x^{l(2^{\frac{2}{2}}+1)}\mid l=2^{\frac{2}{2}}-2=0\}=\{x+1\} \ ({\rm mod} \ f_2(x));$$
$x+1$ is a primitive element of the finite field $\mathcal{F}_3=\frac{\mathbb{F}_2[x]}{\langle f_3(x)\rangle}$ and
\begin{eqnarray*}
\Omega_{3,1}&=&\{x^{-\frac{4}{2}}(x+1)^{l(2^{\frac{4}{2}}+1)}\mid l=0,1,2^{\frac{4}{2}}-2=2\} \ ({\rm mod} \ f_3(x))\\
  &=&\{x^3,x^3+x+1,x+1\}.
\end{eqnarray*}
Moreover, for any $\omega(x)=a+bx+cx^2+dx^3\in \mathcal{F}_3=\frac{\mathbb{F}_2[x]}{\langle f_4(x)\rangle}$ satisfying $(a,b,c,d)
\in \mathbb{F}_2^4\setminus\{(0,0,0,0)\}$, we have
\begin{eqnarray*}
\omega^\prime(x)&=&\delta_4x^{-4}\omega(x^{-1})=x^{11}\omega(x^{14}) \ ({\rm mod} \ f_5(x)=x^4+x^3+1)\\
 &=& (a+b+d)x^3+(a+c+d)x^2+(b+d)x+a+c.
\end{eqnarray*}
Let $\mathcal{K}_j=\frac{\mathbb{F}_2[x]}{\langle f_j(x)^2\rangle}$ for all $j=1,2,3,4,5$. By Corollary 3.5, all $945$ self-dual cyclic codes
over $\mathbb{F}_2+u\mathbb{F}_2$ of length $30$ are given by
$$\mathcal{C}=\mathcal{C}_1\oplus \mathcal{C}_2\oplus \mathcal{C}_3\oplus \mathcal{C}_4,$$
where
\begin{itemize}
\item
  $\mathcal{C}_1=\varepsilon_1(x)C_1$, $C_1$ is one of the following
$3$ ideals of the ring $\mathcal{K}_1+u \mathcal{K}_1$:
$\langle u\rangle$, $\langle (x-1)\rangle$, $\langle u+(x-1)\rangle$.

\vskip 2mm\item
  $\mathcal{C}_2=\varepsilon_2(x)C_2$, $C_2$ is one of the following
$3$ ideals of the ring $\mathcal{K}_2+u \mathcal{K}_2$:
$\langle u\rangle$, $\langle (x^2+x+1)\rangle$, $\langle u+(x^2+x+1)\cdot (x+1)\rangle$.

\vskip 2mm\item
  $\mathcal{C}_3=\varepsilon_3(x)C_3$, $C_3$ is one of the following
$5$ ideals of the ring $\mathcal{K}_3+u \mathcal{K}_3$:
$\langle u\rangle$, $\langle (x^4+x^3+x^2+x+1)\rangle$, $\langle u+(x^4+x^3+x^2+x+1)\cdot \omega(x)\rangle$ with
$\omega(x)\in \Omega_{3,1}$.

\vskip 2mm\item
  $\mathcal{C}_4=\varepsilon_4(x)C_4\oplus \varepsilon_5(x)C_5$, $C_j$
is an ideal of the ring $\mathcal{K}_j+u \mathcal{K}_j$ for $j=4,5$, and the pair $(C_4,C_5)$ is one of the following
$21$ cases:
\begin{description}
\item{}
   $C_4=\langle u^i\rangle$ and $C_5=\langle u^{2-i}\rangle$, where $i=0,1,2$;

\item{}
   $C_4=\langle f_4(x)\rangle$ and $C_5=\langle f_{5}(x)\rangle$;

\item{}
   $C_4=\langle uf_4(x)\rangle$ and $C_5=\langle u, f_{5}(x)\rangle$;

\item{}
   $C_4=\langle u,f_4(x)\rangle$ and $C_5=\langle uf_{5}(x)\rangle$;

\item{}
  $C_4=\langle u+f_4(x)(a+bx+cx^2+dx^3)\rangle$
   and $C_5=\langle u+f_{5}(x)((a+b+d)x^3+(a+c+d)x^2+(b+d)x+a+c)\rangle$,
where $(a,b,c,d)
\in \mathbb{F}_2^4\setminus\{(0,0,0,0)\}$.
\end{description}
\end{itemize}

\par
   Finally, by Lemma 4.2 and Theorem 4.3 we obtain $945$ binary self-dual and
$2$-quasi-cyclic codes $\phi(\mathcal{C})$ of length $60$. For example, among these codes we have the following
$48$ self-dual and $2$-quasi-cyclic codes $\phi(\mathcal{C})$ with basic parameters $[60, 30, 8]$, which are determined by:
\begin{itemize}
\item
$C_2$ is $\langle u\rangle$ or $\langle u+(x^2+x+1)\cdot (x+1)\rangle$.

\item
 The pair $(C_4,C_5)$ is $(\langle u^i\rangle,\langle u^{2-i}\rangle)$, for $i=0,2$.

\item
 The pair $(C_1,C_3)$ is one of the following $12$ cases:
\begin{description}
\item{$\triangleright$}
$C_1=\langle u\rangle$, and $C_3$ is one of the following
$4$ ideals: \\
 $\langle (x^4+x^3+x^2+x+1)\rangle$, $\langle u+(x^4+x^3+x^2+x+1)\cdot \omega(x)\rangle$ with
$\omega(x)\in \Omega_{3,1}$;

\item{$\triangleright$}
 $C_1=\langle (x-1)\rangle$, and $C_3$ is one of the following
$4$ ideals: \\
$\langle u\rangle$, $\langle u+(x^4+x^3+x^2+x+1)\cdot \omega(x)\rangle$ with
$\omega(x)\in \Omega_{3,1}$;

\item{$\triangleright$}
 $C_1=\langle u+(x-1)\rangle$, and $C_3$ is one of the following
$4$ ideals: \\
$\langle u\rangle$, $\langle (x^4+x^3+x^2+x+1)\rangle$, $\langle u+(x^4+x^3+x^2+x+1)\cdot \omega(x)\rangle$ with
$\omega(x)\in \{x^3,x^3+x+1\}$.

\end{description}
\end{itemize}


\section{The hull of every cyclic code
with length $2n$ over $\mathbb{F}_{2^m}+u\mathbb{F}_{2^m}$}

\noindent
 In this section, we determine the hull of each
cyclic code over $\mathbb{F}_{2^m}+u\mathbb{F}_{2^m}$ with length $2n$ where
$n$ is odd.

\par
  As a generalization for the hull of a linear code over finite field, for any linear code
$\mathcal{C}$ of length $2n$ over the ring $\mathbb{F}_{2^m}+u\mathbb{F}_{2^m}$, the hull of
$\mathcal{C}$ is defined by
$${\rm Hull}(\mathcal{C})=\mathcal{C}\cap \mathcal{C}^{\bot}.$$
Let $\phi$ be the isomorphism of $\mathbb{F}_{2^m}$-linear spaces
from $\frac{(\mathbb{F}_{2^m}+u \mathbb{F}_{2^m})[x]}{\langle x^{2n}-1\rangle}$ onto $\mathbb{F}_{2^m}^{4n}$ defined
by Equation (8) in Section 4. Then from properties for ideals in a ring and Lemma 4.2, we deduce the following conclusion immediately.

\vskip 3mm\noindent
  {\bf Proposition 5.1} \textit{Let $\mathcal{C}$ be a cyclic code of length $2n$ over $\mathbb{F}_{2^m}+u\mathbb{F}_{2^m}$
and $\phi(\mathcal{C})$ be defined as in Lemma 4.2. Then}

\begin{description}
\vskip 2mm
\item{(i)}
  \textit{${\rm Hull}(\mathcal{C})$ is a cyclic code  over $\mathbb{F}_{2^m}+u\mathbb{F}_{2^m}$ of length $2n$}.

\vskip 2mm
\item{(ii)}
  \textit{${\rm Hull}(\phi(\mathcal{C}))=\phi(\mathcal{C})\cap (\phi(\mathcal{C}))^{\bot}$ is a $2$-quasi-cyclic code over $\mathbb{F}_{2^m}$ of length $4n$, and ${\rm Hull}(\phi(\mathcal{C}))=\phi({\rm Hull}(\mathcal{C}))$}.

\vskip 2mm
\item{(iii)}
 \textit{The Hamming weight distribution of ${\rm Hull}(\phi(\mathcal{C}))$ is exactly the same as the Lee weight distribution of ${\rm Hull}(\mathcal{C})$}.
\end{description}

\vskip 3mm \par
  It is known that a class of entanglement-assisted quantum
error correcting codes (EAQECCs) can be constructed from
classical codes and their basic parameters are related to the hulls of classical codes
 ([11] Corollary 3.1):
\begin{description}
\item{}
\ \  \textit{Let $\mathcal{C}$ be a classical $[n, k, d]_q$ linear code and $\mathcal{C}^{\bot}$ its Euclidean dual
with parameters $[n, n-k, d^{\bot}]_q$ . Then there exist $[[n, k-{\rm dim}({\rm Hull}(\mathcal{C})), d; n-k-
{\rm dim}({\rm Hull}(\mathcal{C}))]]_q$ and $[[n, n-k-{\rm dim}({\rm Hull}(\mathcal{C})), d^{\bot};
k-{\rm dim}({\rm Hull}(\mathcal{C}))]]_q$ EAQECCs.
Further, if C is MDS then the two EAQECCs are also MDS}.
\end{description}

\par
 Using the notation in the beginning of Section 3, we
denote
\begin{equation}
\varrho(j)=\left\{\begin{array}{ll} j, & {\rm} \ {\rm when} \ 1\leq j\leq \lambda; \cr
                                      j+\epsilon, & {\rm} \ {\rm when} \ \lambda+1\leq j\leq \lambda+\epsilon; \cr
                                      j-\epsilon, & {\rm} \ {\rm when} \ \lambda+\epsilon+1\leq j\leq \lambda+2\epsilon.
                                      \end{array}\right.
\end{equation}
The dual code for every cyclic code of length $2n$ over $\frac{\mathbb{F}_{2^m}[u]}{\langle u^k\rangle}$,
where $k\geq 2$, had been determined by [7] Theorem 3.5.  Here we
restatement the conclusion for the case $k=2$ as follows.

\vskip 3mm \noindent
  {\bf Lemma 5.2} \textit{Let $\mathcal{C}$ be a cyclic code
of length $2n$ over $\mathbb{F}_{2^m}+u\mathbb{F}_{2^m}$ $(u^2=0)$ with canonical form decomposition
$\mathcal{C}=\bigoplus_{j=1}^r\varepsilon_j(x)C_j$, where $C_j$ is an ideal of
$\mathcal{K}_j+u\mathcal{K}_j$. Then}
\begin{description}
\item{$\bullet$}
  \textit{$|\mathcal{C}|=\prod_{j=1}^r|C_j|$ and ${\rm dim}_{\mathbb{F}_{2^m}}(\mathcal{C})=\sum_{j=1}^r{\rm dim}_{\mathbb{F}_{2^m}}(C_j)$}.

\item{$\bullet$}
\textit{The dual code of $\mathcal{C}$ is given by}
$$\mathcal{C}^{\bot}=\bigoplus_{j=1}^r\varepsilon_j(x)D_j,$$
\textit{where $D_j$ is an ideal of
$\mathcal{K}_j+u\mathcal{K}_j$ determined by the following table for all $j=1,\ldots,r$}:
\begin{center}
\begin{tabular}{llll|l}\hline
$\mathcal{L}$ &  $C_j$ (mod $f_j(x)^2$) & $|C_j|$ & $\kappa_j$
& $D_{\varrho(j)}$  (mod $f_{\varrho(j)}(x)^2$)\\ \hline
$1$ & $\bullet$ $\langle 0\rangle$  & $1$ & $0$ & $\diamond$ $\langle 1\rangle$ \\
$1$ & $\bullet$ $\langle 1\rangle$  & $4^{2d_jm}$ & $4d_j$ & $\diamond$ $\langle 0\rangle$ \\
$1$ & $\bullet$ $\langle u\rangle$  & $4^{d_jm}$ & $2d_j$ & $\diamond$ $\langle u\rangle$ \\
$1$ & $\bullet$  $\langle f_j(x)\rangle$ & $4^{d_jm}$ & $2d_j$ & $\diamond$ $\langle f_{\varrho(j)}(x)\rangle$ \\
$1$ & $\bullet$  $\langle uf_j(x)\rangle$ & $2^{d_jm}$ & $d_j$ & $\diamond$ $\langle u,f_{\varrho(j)}(x)\rangle$ \\
$2^{d_jm}-1$ & $\bullet$ $\langle u+f_j(x)\omega\rangle$ & $4^{d_jm}$ & $2d_j$ & $\diamond$ $\langle u+f_{\varrho(j)}(x)\omega^{\prime}\rangle$ \\
$1$ & $\bullet$ $\langle u,f_j(x)\rangle$ & $2^{3d_jm}$ & $3d_j$ & $\diamond$ $\langle uf_{\varrho(j)}(x)\rangle$ \\
 \hline
\end{tabular}
\end{center}
\textit{where} \textit{$\kappa_j={\rm dim}_{\mathbb{F}_{2^m}}(C_j)$, $\mathcal{L}$ is the number of pairs $(C_j,D_{\varrho(j)})$ in the same row, and}
\begin{description}
\item{}
  \textit{$\omega=\omega(x)\in \mathcal{F}_j
  =\{\sum_{i=0}^{d_j-1}a_ix^i\mid a_0,a_1,\ldots,a_{d_j-1}\in \mathbb{F}_{2^m}\}$ and $\omega\neq 0$},

\item{}
 $\omega^{\prime}=\delta_jx^{-d_j}\omega(x^{-1})$ $({\rm mod} \ f_{\varrho(j)}(x))$.
\end{description}
\end{description}

\vskip 3mm\par
  For each integer $j$, $1\leq j\leq r$, let $\mathcal{F}_j\setminus\{0\}=\{\omega_1,\ldots,\omega_{2^{d_jm}-1}\}$. Then
  the ideal lattice of the ring
$\mathcal{K}_j+u\mathcal{K}_j$ is the following figure.
$$
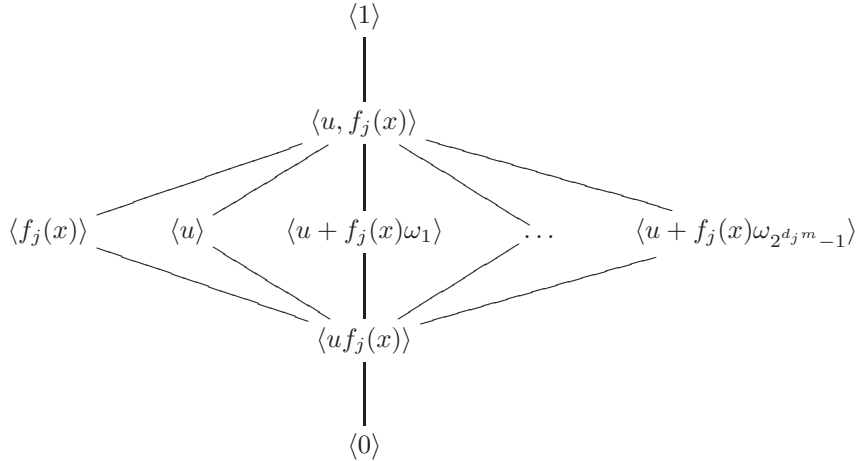
\xymatrix{
&  & \langle 1\rangle \ar@{-}[d] & & \\
& & \langle u,f_j(x)\rangle \ar@{-}[dll] \ar@{-}[dl] \ar@{-}[d] \ar@{-}[dr] \ar@{-}[drr]& &\\
 \langle f_j(x)\rangle & \langle u\rangle & \langle u+f_j(x)\omega_1\rangle & \ldots & \langle u+f_j(x)\omega_{2^{d_jm}-1}\rangle \\
&   & \langle u f_j(x)\rangle \ar@{-}[ull]  \ar@{-}[ul] \ar@{-}[u] \ar@{-}[ur] \ar@{-}[urr]& &\\
&  & \langle 0\rangle \ar@{-}[u] & & }$$
Then we have the following conclusion.

\vskip 3mm\noindent
  {\bf Theorem 5.3} \textit{Let $\mathcal{C}$ be a cyclic code
of length $2n$ over $\mathbb{F}_{2^m}+u\mathbb{F}_{2^m}$ with canonical form decomposition
$\mathcal{C}=\bigoplus_{j=1}^r\varepsilon_j(x)C_j$, where $C_j$ is an ideal of
$\mathcal{K}_j+u\mathcal{K}_j$. Then the Hull of $\mathcal{C}$ is given by}
$${\rm Hull}(\mathcal{C})=\bigoplus_{j=1}^r\varepsilon_j(x)H_j,$$
\textit{where $H_j$ is an ideal of
$\mathcal{K}_j+u\mathcal{K}_j$ determined by the following conditions for all $j=1,\ldots,r$}:
\begin{description}
\item{(i)}
 \textit{Let $j=1$. Then}
$$H_1=\left\{\begin{array}{ll} \langle 0\rangle, & {\rm if} \
C_1=\langle 0\rangle \ {\rm or} \ \langle 1\rangle; \cr
\langle u(x-1)\rangle, & {\rm if} \
C_1=\langle u(x-1)\rangle \ {\rm or} \ \langle u, x-1\rangle; \cr
\langle x-1\rangle, & {\rm if} \
C_1=\langle x-1\rangle; \cr
\langle u+(x-1)a\rangle, & {\rm if} \
C_1=\langle u+(x-1)a\rangle \ {\rm where} \ a\in \mathbb{F}_{2^m}.
\end{array}\right.$$

\item{(ii)}
   \textit{Let $2\leq j\leq \lambda$. Then}
$$H_j=\left\{\begin{array}{ll} \langle 0\rangle, & {\rm if} \
C_j=\langle 0\rangle \ {\rm or} \ \langle 1\rangle; \cr
\langle uf_j(x)\rangle, & {\rm if} \
C_j=\langle uf_j(x)\rangle \ {\rm or} \ \langle u, f_j(x)\rangle; \cr
\langle f_j(x)\rangle, & {\rm if} \
C_j=\langle f_j(x)\rangle; \cr
\langle u+f_j(x)\omega\rangle, & {\rm if} \
C_j=\langle u+f_j(x)\omega\rangle \ {\rm where} \ \omega\in \{0\}\cup \Theta_{j,1}; \cr
\langle uf_j(x)\rangle, & {\rm if} \
C_j=\langle u+f_j(x)\omega\rangle \ {\rm where} \ 0\neq\omega\in \mathcal{F}_j\setminus\Theta_{j,1}.
\end{array}\right.$$

\item{(iii)}
 \textit{Let $\lambda+1\leq j\leq \lambda+\epsilon$. Then the pair $(H_j,H_{j+\epsilon})$ of
 ideals is given by one of the following six cases, where
 $\mathcal{S}_{j+\epsilon}$ is the set of all $5+2^{d_jm}$ ideals in the ring $\mathcal{K}_{j+\epsilon}+u\mathcal{K}_{j+\epsilon}$
 listed by Lemma 5.2}.

\begin{description}
\item{{\bf 1}.}
  \textit{Let $C_j=\langle 0\rangle$. Then}
 \begin{description}
\item{$\diamond$}
  \textit{$H_j=\langle 0\rangle$ and $H_{j+\epsilon}=C_{j+\epsilon}$, for every $C_{j+\epsilon}\in \mathcal{S}_{j+\epsilon}$}.
 \end{description}

\vskip 2mm
\item{{\bf 2}.}
  \textit{Let $C_j=\langle uf_j(x)\rangle$. Then}
 \begin{description}
\item{$\diamond$}
  \textit{$H_j=\langle 0\rangle$ and $H_{j+\epsilon}=\langle u, f_{j+\epsilon}(x)\rangle$, if $C_{j+\epsilon}=\langle 1\rangle$};

\item{$\diamond$}
  \textit{$H_j=\langle uf_j(x)\rangle$ and $H_{j+\epsilon}=C_{j+\epsilon}$, if $C_{j+\epsilon}\in \mathcal{S}_{j+\epsilon}$ and $C_{j+\epsilon}\neq \langle 1\rangle$}.
 \end{description}

\vskip 2mm
\item{{\bf 3}.}
  \textit{Let $C_j=\langle f_j(x)\rangle$. Then}
 \begin{description}
\item{$\diamond$}
  \textit{$H_j=\langle f_j(x)\rangle$ and $H_{j+\epsilon}=\langle f_{j+\epsilon}(x)\rangle$, if $C_{j+\epsilon}=\langle f_{j+\epsilon}(x)\rangle$};

\item{$\diamond$}
 \textit{$H_j=\langle uf_j(x)\rangle$ and $H_{j+\epsilon}=\langle f_{j+\epsilon}(x)\rangle$,
 if $C_{j+\epsilon}=\langle u, f_{j+\epsilon}(x)\rangle$};

\item{$\diamond$}
 \textit{$H_j=\langle 0\rangle$ and $H_{j+\epsilon}=\langle f_{j+\epsilon}(x)\rangle$, if $C_{j+\epsilon}=\langle 1\rangle$};

\item{$\diamond$}
 \textit{$H_j=\langle f_j(x)\rangle$ and $H_{j+\epsilon}=C_{j+\epsilon}$, if $C_{j+\epsilon}=\langle uf_{j+\epsilon}(x)\rangle$,  $\langle 0\rangle$};

\item{$\diamond$}
 \textit{$H_j=\langle uf_j(x)\rangle$ and $H_{j+\epsilon}=\langle uf_{j+\epsilon}(x)\rangle$, \\
 if $C_{j+\epsilon}=\langle u+f_{j+\epsilon}(x)\omega^{\prime}\rangle$ for any $\omega^{\prime}\in \mathcal{F}_{j+\epsilon}$}.
 \end{description}

\vskip 2mm
\item{{\bf 4}.}
  \textit{Let $C_j=\langle u+f_j(x)\omega_0\rangle$, where
 $\omega_0=\omega_0(x)\in\mathcal{F}_j$. Denote $\omega_0^\prime=\omega_0^\prime(x)=\delta_jx^{-d_j}\omega_0(x^{-1})$
 $({\rm mod} \ f_{j+\epsilon}(x))$ in the following. Especially, we have
 $\omega_0^\prime=0$ when $\omega_0=0$. Then}

\begin{description}
\item{$\diamond$}
  \textit{$H_j=\langle u+f_j(x)\omega_0\rangle$ and $H_{j+\epsilon}=C_{j+\epsilon}$, \\
  if $C_{j+\epsilon}=\langle uf_{j+\epsilon}(x)\rangle$, $\langle u+f_{j+\epsilon}(x)\omega_0^\prime \rangle$, $\langle 0\rangle$};

\item{$\diamond$}
 \textit{$H_j=\langle uf_{j}(x)\rangle$ and $H_{j+\epsilon}=\langle u+f_{j+\epsilon}(x)\omega_0^\prime\rangle$,
 if $C_{j+\epsilon}=\langle u, f_{j+\epsilon}(x)\rangle$};

\item{$\diamond$}
 \textit{$H_j=\langle 0\rangle$ and $H_{j+\epsilon}=\langle u+f_{j+\epsilon}(x)\omega_0^\prime\rangle$,
 if $C_{j+\epsilon}=\langle 1\rangle$};

\item{$\diamond$}
 \textit{$H_j=\langle uf_j(x)\rangle$ and $H_{j+\epsilon}=\langle uf_{j+\epsilon}(x)\rangle$, \\
 if $C_{j+\epsilon}=\langle f_{j+\epsilon}(x)\rangle$,
 $\langle u+f_{j+\epsilon}(x)\omega^\prime\rangle$
 where $\omega^\prime\in \mathcal{F}_{j+\epsilon}\setminus\{\omega_0^\prime\}$}.
 \end{description}

\vskip 2mm
\item{{\bf 5}.}
  \textit{Let $C_j=\langle u, f_j(x)\rangle$. Then}
\begin{description}
\item{$\diamond$}
  \textit{$H_j=\langle u, f_j(x)\rangle$ and $H_{j+\epsilon}=\langle 0\rangle$,
  if $C_{j+\epsilon}=\langle 0\rangle$};

\item{$\diamond$}
  \textit{$H_j=D_{\varrho(j+\epsilon)}$ and $H_{j+\epsilon}=\langle uf_{j+\epsilon}(x)\rangle$,
  if $C_{j+\epsilon}\in \mathcal{S}_{j+\epsilon}\setminus\{\langle 0\rangle\}$}.
\end{description}

\vskip 2mm
\item{{\bf 6}.}
  \textit{Let $C_j=\langle 1\rangle$. Then}
\begin{description}
\item{$\diamond$}
 \textit{$H_j=D_{\varrho(j+\epsilon)}$ and $H_{j+\epsilon}=\langle 0\rangle$, for every $C_{j+\epsilon}\in \mathcal{S}_{j+\epsilon}$}.
\end{description}
\end{description}
\end{description}

\par
  \textit{Moreover, we have that ${\rm dim}_{\mathbb{F}_{2^m}}({\rm Hull}(\mathcal{C}))
=\sum_{j=1}^r{\rm dim}_{\mathbb{F}_{2^m}}(H_j)$}.

\vskip 3mm \noindent
 \textit{Remark} ($\dag$) In Cases 5 and 6 of (iii) above, by Lemma 5.2
and $\varrho(j+\epsilon)=j$ the ideal
$D_{\varrho(j+\epsilon)}$ in the ring $\mathcal{K}_j+u\mathcal{K}_j$ is determined by
the following table:
\begin{center}
\begin{tabular}{lll|l}\hline
$\mathcal{L}$ &  $C_{j+\epsilon}$ (mod $f_{j+\epsilon}(x)^2$) & $|C_{j+\epsilon}|$ & $D_{\varrho(j)}=D_j$  (mod $f_{j}(x)^2$)\\ \hline
$1$ & $\bullet$ $\langle 0\rangle$  & $1$ & $\diamond$ $\langle 1\rangle$ \\
$1$ & $\bullet$ $\langle 1\rangle$  & $4^{2d_jm}$ & $\diamond$ $\langle 0\rangle$ \\
$1$ & $\bullet$ $\langle u\rangle$  & $4^{d_jm}$ & $\diamond$ $\langle u\rangle$ \\
$1$ & $\bullet$  $\langle f_{j+\epsilon}(x)\rangle$ & $4^{d_jm}$ & $\diamond$ $\langle f_{j}(x)\rangle$ \\
$1$ & $\bullet$  $\langle uf_{j+\epsilon}(x)\rangle$ & $2^{d_jm}$ & $\diamond$ $\langle u,f_{j}(x)\rangle$ \\
$2^{d_jm}-1$ & $\bullet$ $\langle u+f_{j+\epsilon}(x)\omega\rangle$ & $4^{d_jm}$ & $\diamond$ $\langle u+f_{j}(x)\omega^{\prime}\rangle$ \\
$1$ & $\bullet$ $\langle u,f_{j+\epsilon}(x)\rangle$ & $2^{3d_jm}$ & $\diamond$ $\langle uf_{j}(x)\rangle$ \\
 \hline
\end{tabular}
\end{center}
\textit{where} \textit{$\mathcal{L}$ is the number of pairs $(C_{j+\epsilon},D_j)$ in the same row, $d_j=d_{j+\epsilon}$ and}
\begin{description}
\item{}
  \textit{$\omega=\omega(x)\in \mathcal{F}_{j+\epsilon}
  =\{\sum_{i=0}^{d_j-1}a_ix^i\mid a_0,a_1,\ldots,a_{d_j-1}\in \mathbb{F}_{2^m}\}$ and $\omega\neq 0$},

\item{}
 $\omega^{\prime}=\delta_{j+\epsilon}x^{-d_j}\omega(x^{-1})$ $({\rm mod} \ f_{j}(x))$.
\end{description}

\par
  ($\ddag$) The $\mathbb{F}_{2^m}$-dimension ${\rm dim}_{\mathbb{F}_{2^m}}(H_j)$
can be obtained easily through the table in Lemma 5.2.

\vskip 3mm \noindent
 \textit{Proof} Let $\mathcal{C}^{\bot}=\bigoplus_{j=1}^r\varepsilon_j(x)D_j$, where
$D_j$ is an ideal of the ring $\mathcal{K}_j+u\mathcal{K}_j$ determined by
Lemma 5.2 for $j=1,\ldots,r$. Since
$\varepsilon_j(x)^2=\varepsilon_j(x)$ and $\varepsilon_j(x)\varepsilon_l(x)=0$ in the ring
$\mathcal{A}$ for all $j\neq l$ and $j,l=1,\ldots,r$, by Lemma 2.1 it follows that
${\rm Hull}(\mathcal{C})=\mathcal{C}\cap \mathcal{C}^{\bot}=\bigoplus_{j=1}^r\varepsilon_j(x)H_j$ where
$H_j=C_j\cap D_j$ for all $j=1,\ldots,r$. Then by Lemma 5.2 we have the following three cases.

\par
  {\bf Case i}: $j=1$. In this case, we have $\varrho(1)=1$, $f_1(x)=x-1$ and $\mathcal{F}_1=\mathbb{F}_{2^m}$.
By Lemma 3.2, we know that $\delta_1=1$, $d_1=1$ and $\omega^\prime=\omega$ for
any $\omega\in \mathbb{F}_{2^m}\setminus\{0\}$.
  Then we have one of the following
four subcases:

\par
  (i-1) Let $C_1=\langle 0\rangle$ or $\langle 1\rangle$. Then $H_1=C_1\cap D_1=\langle 0\rangle$.

\par
  (i-2) Let $C_1=\langle u(x-1)\rangle$ or $\langle u,x-1\rangle$. Then $H_1=C_1\cap D_1=\langle u(x-1)\rangle$.

\par
  (i-3) Let $C_1=\langle x-1\rangle$. Then $H_1=C_1\cap D_1=\langle x-1\rangle$.
\par
  (i-4) Let $C_1=\langle u+(x-1)a\rangle$, where
$a\in \mathcal{F}_{2^m}$. Then $D_1=\langle u+(x-1)a\rangle$ by Lemma 5.2. This implies
$H_1=C_1\cap D_1=\langle u+(x-1)a\rangle$.

\par
  {\bf Case ii}: $2\leq j\leq \lambda$. In this case, we have $\varrho(j)=j$,
$\mathcal{F}_j=\frac{\mathbb{F}_{2^m}[x]}{\langle f_j(x)\rangle}$.
By Lemma 3.2, we know that $\delta_j=1$.
  Then we have one of the following
five subcases:

\par
  (ii-1) Let $C_j=\langle 0\rangle$ or $\langle 1\rangle$. Then $H_j=C_j\cap D_j=\langle 0\rangle$.

\par
  (ii-2) Let $C_j=\langle uf_j(x)\rangle$ or $\langle u,f_j(x)\rangle$. Then $H_j=C_j\cap D_j=\langle uf_j(x)\rangle$.

\par
  (ii-3) Let $C_j=\langle f_j(x)\rangle$. Then $H_j=C_j\cap D_j=\langle f_j(x)\rangle$.

\par
  (ii-4) Let $C_j=\langle u+f_j(x)\omega\rangle$, where
$\omega=\omega(x)\in \{0\}\cup \Theta_{j,1}$. For any
$\omega\in \Theta_{j,1}$, by the definition of the set
$\Theta_{j,1}$ before Theorem 3.1 we have that
$\omega=\delta_jx^{-d_j}\widehat{\omega}=\omega^{\prime}$ in the finite field $\mathcal{F}_j$. This
implies $D_j=\langle u+f_j(x)\omega\rangle$ for all $\omega\in \{0\}\cup \Theta_{j,1}$.
Hence $H_j=C_j\cap D_j=\langle u+f_j(x)\omega\rangle$.

\par
  (ii-5) Let $C_j=\langle u+f_j(x)\omega\rangle$, where
$\omega\neq 0$ and $\omega\in \mathcal{F}_j\setminus\Theta_{j,1}$. By Lemma 5.2, we have
that $D_j=\langle u+f_j(x)\omega^\prime\rangle$ and $\omega\neq \omega^\prime$. From this we deduce
that $H_j=C_j\cap D_j=\langle uf_j(x)\rangle$.

\par
  {\bf Case iii}: $\lambda+1\leq j\leq \lambda+\epsilon$. In this case, we have $\varrho(j)=j+\epsilon$,
$\varrho(j+\epsilon)=j$,
$\mathcal{F}_j=\frac{\mathbb{F}_{2^m}[x]}{\langle f_j(x)\rangle}$ and
$\mathcal{F}_{j+\epsilon}=\frac{\mathbb{F}_{2^m}[x]}{\langle f_{j+\epsilon}(x)\rangle}$.
Then we have one of the following seven subcases:

\par
  ({\bf iii-1}) $C_j=\langle 0\rangle$. In this case, by Lemma 5.2 we have $D_{j+\epsilon}=\langle 1\rangle$. This implies
$H_j=C_j\cap D_j=\langle 0\rangle$ and $H_{j+\epsilon}=C_{j+\epsilon}\cap D_{j+\epsilon}=C_{j+\epsilon}$ for
any ideal $C_{j+\epsilon}$ of $\mathcal{K}_{j+\epsilon}+u \mathcal{K}_{j+\epsilon}$ by Lemma 5.2.

\par
  ({\bf iii-2}) $C_j=\langle uf_j(x)\rangle$. In this case, we have $D_{j+\epsilon}=\langle u, f_{j+\epsilon}(x)\rangle$. Then by
Lemma 5.2 we have the following conclusions:
\begin{description}
\item{$\triangleright$}
  If $C_{j+\epsilon}=\langle 1\rangle$, then $D_j=D_{\varrho(j+\epsilon)}=\langle 0\rangle$ by Lemma 5.2.
  Hence $H_j=C_j\cap D_j=\langle 0\rangle$ and $H_{j+\epsilon}=C_{j+\epsilon}\cap D_{j+\epsilon}=D_{j+\epsilon}=\langle u, f_{j+\epsilon}(x)\rangle$.

\item{$\triangleright$}
  If $C_{j+\epsilon}\neq \langle 1\rangle$, we have
$C_{j+\epsilon}\subseteq \langle u, f_{j+\epsilon}(x)\rangle$, and that
$D_j=D_{\varrho(j+\epsilon)}\supseteq \langle uf_j(x)\rangle$ by Lemma 5.2.
 Hence $H_j=C_j\cap D_j=C_j=\langle uf_j(x)\rangle$, and
 $H_{j+\epsilon}=C_{j+\epsilon}\cap D_{j+\epsilon}=C_{j+\epsilon}$ for
 any ideal $C_{j+\epsilon}$ of $\mathcal{K}_{j+\epsilon}+u \mathcal{K}_{j+\epsilon}$
 satisfying $C_{j+\epsilon}\neq \langle 1\rangle$.
\end{description}

\par
  ({\bf iii-3}) $C_j=\langle f_j(x)\rangle$. In this case, we have $D_{j+\epsilon}=\langle f_{j+\epsilon}(x)\rangle$. Then by
Lemma 5.2 we have the following conclusions:
\begin{description}
\item{$\triangleright$}
  If $C_{j+\epsilon}=\langle f_{j+\epsilon}(x)\rangle$, $\langle u, f_{j+\epsilon}(x)\rangle$ or $\langle 1\rangle$, then
  $D_j=D_{\varrho(j+\epsilon)}=\langle f_{j}(x)\rangle$, $\langle uf_{j}(x)\rangle$ or $\langle 0\rangle$ respectively.
  Hence $H_j=C_j\cap D_j=D_j$ and $H_{j+\epsilon}=C_{j+\epsilon}\cap D_{j+\epsilon}=D_{j+\epsilon}=\langle  f_{j+\epsilon}(x)\rangle$.

\item{$\triangleright$}
  If $C_{j+\epsilon}=\langle uf_{j+\epsilon}(x)\rangle$ or $\langle 0\rangle$, then
  $D_j=D_{\varrho(j+\epsilon)}=\langle u, f_{j}(x)\rangle$ or $\langle 1\rangle$ respectively.
  Hence $H_j=C_j\cap D_j=C_j=\langle f_j(x) \rangle$ and $H_{j+\epsilon}=C_{j+\epsilon}\cap D_{j+\epsilon}=C_{j+\epsilon}$.

\item{$\triangleright$}
  If $C_{j+\epsilon}=\langle u+f_{j+\epsilon}(x)\omega^\prime\rangle$ where $\omega^\prime=
  \omega^\prime(x)\in \mathcal{F}_{j+\epsilon}$, then
  $D_j=D_{\varrho(j+\epsilon)}=\langle u+f_{j}(x)\omega\rangle$ where $\omega=\omega(x)\in \mathcal{F}_{j}$ satisfying
  $$\omega^\prime(x)=\delta_jx^{-d_j}\omega(x^{-1}) \ ({\rm mod} \ f_{j+\epsilon}(x)).$$
    Hence $H_j=C_j\cap D_j=\langle f_j(x) \rangle\cap \langle u+f_{j}(x)\omega\rangle=\langle uf_j(x)\rangle$ and $H_{j+\epsilon}=C_{j+\epsilon}\cap D_{j+\epsilon}=\langle u+f_{j+\epsilon}(x)\omega^\prime\rangle\cap \langle f_{j+\epsilon}(x)\rangle=\langle uf_{j+\epsilon}(x)\rangle$.
\end{description}

\par
  ({\bf iii-4}) $C_j=\langle u\rangle$. In this case, we have $D_{j+\epsilon}=\langle u\rangle$. Similar to the case (iii-3), by Lemma 5.2 we have the following conclusions:
\begin{description}
\item{$\triangleright$}
  If $C_{j+\epsilon}=\langle u\rangle$, $\langle u, f_{j+\epsilon}(x)\rangle$ or $\langle 1\rangle$, then
  $D_j=D_{\varrho(j+\epsilon)}=\langle u\rangle$, $\langle uf_{j}(x)\rangle$ or $\langle 0\rangle$ respectively.
  Hence $H_j=C_j\cap D_j=D_j$ and $H_{j+\epsilon}=D_{j+\epsilon}=\langle u\rangle$.

\item{$\triangleright$}
  If $C_{j+\epsilon}=\langle uf_{j+\epsilon}(x)\rangle$ or $\langle 0\rangle$, then
  $D_j=D_{\varrho(j+\epsilon)}=\langle u, f_{j}(x)\rangle$ or $\langle 1\rangle$ respectively.
  Hence $H_j=C_j\cap D_j=C_j=\langle u \rangle$ and $H_{j+\epsilon}=C_{j+\epsilon}\cap D_{j+\epsilon}=C_{j+\epsilon}$.

\item{$\triangleright$}
  If $C_{j+\epsilon}=\langle f_{j+\epsilon}(x)\rangle$ or $\langle u+f_{j+\epsilon}(x)\omega^\prime\rangle$ where $\omega^\prime=
  \omega^\prime(x)\in \mathcal{F}_{j+\epsilon}\setminus\{0\}$, then
  $D_j=D_{\varrho(j+\epsilon)}=\langle f_j(x)\rangle$ or $\langle u+f_{j}(x)\omega\rangle$ where
  $\omega=\omega(x)\in \mathcal{F}_{j}\setminus\{0\}$ satisfying
  $\omega^\prime(x)=\delta_jx^{-d_j}\omega(x^{-1}) \ ({\rm mod} \ f_{j+\epsilon}(x)).$
  Hence $H_j=C_j\cap D_j=\langle u \rangle\cap D_j=\langle uf_j(x)\rangle$ and $H_{j+\epsilon}=C_{j+\epsilon}\cap D_{j+\epsilon}=C_{j+\epsilon}\cap \langle u\rangle=\langle uf_{j+\epsilon}(x)\rangle$.
\end{description}

\par
  ({\bf iii-5}) $C_j=\langle u+f_j(x)\omega_0\rangle$
where $\omega_0=\omega_0(x)\in \mathcal{F}_j\setminus\{0\}$. In this case, we have $D_{j+\epsilon}=\langle u+f_{j+\epsilon}(x)\omega_0^\prime\rangle$ where
$\omega_0^\prime\in \mathcal{F}_{j+\epsilon}\setminus\{0\}$ satisfying
$\omega_0^\prime=\omega_0^\prime(x)=\delta_jx^{-d_j}\omega_0(x^{-1})$ (mod $f_{j+\epsilon}(x)$). Then by
Lemma 5.2 we have the following conclusions:
\begin{description}
\item{$\triangleright$}
  If $C_{j+\epsilon}=\langle u+f_{j+\epsilon}(x)\omega_0^\prime\rangle$, $\langle u, f_{j+\epsilon}(x)\rangle$ or $\langle 1\rangle$, then
  $D_j=D_{\varrho(j+\epsilon)}=\langle u+f_j(x)\omega_0\rangle$, $\langle uf_{j}(x)\rangle$ or $\langle 0\rangle$ respectively.
  Hence $H_j=C_j\cap D_j=D_j$ and $H_{j+\epsilon}=C_{j+\epsilon}\cap D_{j+\epsilon}=C_{j+\epsilon}
  =\langle  u+f_{j+\epsilon}(x)\omega_0^\prime\rangle$.

\item{$\triangleright$}
  If $C_{j+\epsilon}=\langle uf_{j+\epsilon}(x)\rangle$ or $\langle 0\rangle$, then
  $D_j=D_{\varrho(j+\epsilon)}=\langle u, f_{j}(x)\rangle$ or $\langle 1\rangle$ respectively.
  Hence $H_j=C_j=\langle u+f_j(x)\omega_0 \rangle$ and $H_{j+\epsilon}=C_{j+\epsilon}\cap D_{j+\epsilon}=C_{j+\epsilon}$.

\item{$\triangleright$}
  If $C_{j+\epsilon}=\langle f_{j+\epsilon}(x)\rangle, \langle u\rangle$ or $\langle u+f_{j+\epsilon}(x)\omega^\prime\rangle$ where $\omega^\prime=
  \omega^\prime(x)\in \mathcal{F}_{j+\epsilon}\setminus\{\omega_0^{\prime}\}$, then
  $D_j=D_{\varrho(j+\epsilon)}=\langle f_j(x)\rangle, \langle u\rangle$ or $\langle u+f_{j}(x)\omega\rangle$ where
  $\omega=\omega(x)\in \mathcal{F}_{j}\setminus\{\omega_0\}$ satisfying
  $\omega^\prime(x)=\delta_jx^{-d_j}\omega(x^{-1}) \ ({\rm mod} \ f_{j+\epsilon}(x)).$
    Hence $H_j=C_j\cap D_j=\langle u+f_j(x)\omega_0 \rangle\cap D_j=\langle uf_j(x)\rangle$ and $H_{j+\epsilon}=C_{j+\epsilon}\cap D_{j+\epsilon}=C_{j+\epsilon}\cap \langle u+f_{j+\epsilon}(x)\omega_0^\prime\rangle=\langle uf_{j+\epsilon}(x)\rangle$.
\end{description}

\par
   ({\bf iii-6}) $C_j=\langle u, f_j(x)\rangle$. In this case, we have $D_{j+\epsilon}=\langle u f_{j+\epsilon}(x)\rangle$. Then by
Lemma 5.2 we have the following conclusions:
\begin{description}
\item{$\triangleright$}
  If $C_{j+\epsilon}=\langle 0\rangle$, then $D_j=D_{\varrho(j+\epsilon)}=\langle 1\rangle$ by Lemma 5.2.
  Hence $H_j=C_j\cap D_j=C_j=\langle u, f_j(x)\rangle$ and $H_{j+\epsilon}=C_{j+\epsilon}\cap D_{j+\epsilon}
  =\langle 0\rangle$.

\item{$\triangleright$}
  If $C_{j+\epsilon}\neq \langle 0\rangle$, we have
$C_{j+\epsilon}\supseteq \langle uf_{j+\epsilon}(x)\rangle$, and that
$D_j=D_{\varrho(j+\epsilon)}\subseteq \langle u, f_j(x)\rangle$ by Lemma 5.2.
 Hence $H_j=C_j\cap D_j=D_j$, and
 $H_{j+\epsilon}=C_{j+\epsilon}\cap D_{j+\epsilon}=C_{j+\epsilon}=\langle uf_{j+\epsilon}(x)\rangle$ for
 any ideal $C_{j+\epsilon}$ of $\mathcal{K}_{j+\epsilon}+u \mathcal{K}_{j+\epsilon}$
 satisfying $C_{j+\epsilon}\neq \langle 0\rangle$.
\end{description}

\par
  ({\bf iii-7}) $C_j=\langle 1\rangle$. In this case, we have $D_{j+\epsilon}=\langle 0\rangle$. This implies
$H_j=C_j\cap D_j=D_j=D_{\varrho(j+\epsilon)}$ and $H_{j+\epsilon}=C_{j+\epsilon}\cap D_{j+\epsilon}=\langle 0\rangle$ for
any ideal $C_{j+\epsilon}$ of $\mathcal{K}_{j+\epsilon}+u \mathcal{K}_{j+\epsilon}$ by Lemma 5.2.

\par
  When $\omega_0=\omega_0(x)=0$, we have $\omega_0^\prime=\omega_0^\prime(x)=\delta_jx^{-d_j}\omega_0(x^{-1})=0$ as well.
Hence the Case (iii-4) and Case (iii-5) can be combined into one case where $\omega_0\in \mathcal{F}_j$.
\hfill $\Box$

\vskip 3mm\par
   For any cyclic code $\mathcal{C}$ of length $2n$ over $\mathbb{F}_{2^m}+u\mathbb{F}_{2^m}$, it is
clear that $\mathcal{C}$ is self-orthogonal if and only if $\mathcal{C}\subseteq \mathcal{C}^{\bot}$. The latter
is equivalent to that ${\rm Hull}(\mathcal{C})=\mathcal{C}$. From this and by Theorem 5.3, we deduce
the following corollary immediately.

\vskip 3mm\noindent
  {\bf Corollary 5.4} \textit{Using the notation in Theorem 5.3, all distinct self-orthogonal cyclic codes
 of length $2n$ over $\mathbb{F}_{2^m}+u\mathbb{F}_{2^m}$ are given by}
$$\mathcal{C}=\bigoplus_{j=1}^r\varepsilon_j(x)C_j \ ({\rm mod} \ x^{2n}-1),$$
\textit{where $C_j$ is an ideal of the ring $\mathcal{K}_j+u\mathcal{K}_j$ listed as follows}.
\begin{description}
\item{(i)}
  \textit{$C_1$ is one of the following $3+2^m$ ideals}:

  \textit{$\langle 0\rangle$, $\langle u(x-1)\rangle$, $\langle x-1\rangle$,
  $\langle u+(x-1)a\rangle$ where $a\in \mathbb{F}_{2^m}$}.

\vskip 2mm\item{(ii)}
  \textit{Let $2\leq j\leq \lambda$. Then $C_j$ is one of the following $3+2^{\frac{d_j}{2}m}$ ideals}:

  \textit{$\langle 0\rangle$, $\langle uf_j(x)\rangle$, $\langle f_j(x)\rangle$,
  $\langle u+f_j(x)\omega\rangle$ where $\omega\in \{0\}\cup \Theta_{j,1}$}.

\vskip 2mm\item{(iii)}
   \textit{Let $\lambda+1\leq j\leq \lambda+\epsilon$. Then the pair
$(C_j,C_{j+\epsilon})$ of ideals is given by one of the following five subcases}:

\begin{description}
\item{$\diamond$}
  \textit{$5+2^{d_jm}$ pairs}:
  $\left\{\begin{array}{l} C_j=\langle 0\rangle, \cr C_{j+\epsilon}\in \mathcal{S}_{j+\epsilon}.\end{array}\right.$

\item{$\diamond$}
  \textit{$4+2^{d_jm}$ pairs}:
  $\left\{\begin{array}{l} C_j=\langle uf_j(x)\rangle, \cr C_{j+\epsilon}\in \mathcal{S}_{j+\epsilon}
  \ {\rm and} \ C_{j+\epsilon}\neq \langle 1\rangle.\end{array}\right.$

\item{$\diamond$}
  \textit{$3$ pairs}:
  $\left\{\begin{array}{l} C_j=\langle f_j(x)\rangle, \cr
  C_{j+\epsilon}=\langle f_{j+\epsilon}(x)\rangle, \ \langle uf_{j+\epsilon}(x)\rangle, \ \langle 0\rangle.\end{array}\right.$

\item{$\diamond$}
  \textit{$3\cdot 2^{d_jm}$ pairs}:
  $\left\{\begin{array}{l} C_j=\langle u+f_j(x)\omega_0\rangle,  \cr
  C_{j+\epsilon}=\langle u+f_{j+\epsilon}(x)\omega_0^\prime\rangle, \ \langle uf_{j+\epsilon}(x)\rangle, \  \langle 0\rangle;\end{array}\right.$ $\forall \omega_0\in \mathcal{F}_j$.

\item{$\diamond$}
  \textit{$2$ pairs}:
  $\left\{\begin{array}{l} C_j=\langle u, f_j(x)\rangle, \ \langle 1\rangle\cr
  C_{j+\epsilon}=\langle 0\rangle.\end{array}\right.$
\end{description}
\end{description}

\par
  \textit{Therefore, the number of self-orthogonal cyclic codes
 of length $2n$ over $\mathbb{F}_{2^m}+u\mathbb{F}_{2^m}$ is}
 $$(3+2^m)\cdot \prod_{j=2}^\lambda(3+2^{\frac{d_j}{2}m})
 \cdot \prod_{j=\lambda+1}^{\lambda+\epsilon}(14+5\cdot 2^{d_jm}).$$

\vskip 3mm \noindent
 \textit{Remark} ($\dag$) Let $\phi$ be the isomorphism of $\mathbb{F}_{2^m}$-linear spaces
from $\frac{(\mathbb{F}_{2^m}+u \mathbb{F}_{2^m})[x]}{\langle x^{2n}-1\rangle}$ onto $\mathbb{F}_{2^m}^{4n}$ defined
by Equation (8) in Section 4. By Proposition 5.1, we see that
$\phi(\mathcal{C})$ is a self-orthogonal $2$-quasi-cyclic code of length $4n$
over the finite field $\mathbb{F}_{2^m}$ for every self-orthogonal cyclic code $\mathcal{C}$
over the ring $\mathbb{F}_{2^m}+u \mathbb{F}_{2^m}$ of length $2n$. In particular,
The Hamming weight distribution of $\phi(\mathcal{C})$ is the same as the Lee weight distribution of $\mathcal{C}$
by Lemma 4.2(ii).

\par
  ($\ddag$) For any cyclic code $\mathcal{C}$ of length $2n$ over $\mathbb{F}_{2^m}+u\mathbb{F}_{2^m}$, recall that
$\mathcal{C}$ is said to be \textit{orthogonal self-contained} if $\mathcal{C}^{\bot}\subseteq \mathcal{C}$.
The latter condition
is equivalent to that ${\rm Hull}(\mathcal{C})=\mathcal{C}^{\bot}$.

  All orthogonal self-contained cyclic codes
of length $2n$ over $\mathbb{F}_{2^m}+u\mathbb{F}_{2^m}$ can be determined by Theorem 5.3. From these codes,
we can obtain orthogonal self-contained and $2$-quasi-cyclic codes
of length $4n$ over $\mathbb{F}_{2^m}$ by Equation (8) in Section 4.
A class of EAQECCs had been constructed from orthogonal
self-contained cyclic codes in many literatures (cf. [11] Proposition 4.2).

\vskip 3mm\par
  Finally, we list the number $\mathcal{NO}$ of self-orthogonal cyclic codes
$\mathcal{C}$ over $\mathbb{F}_2+u\mathbb{F}_2$ of length $2n$, where
$n$ is odd and $6 \leq 2n \leq 98$, in the following table.
\begin{center}
{\footnotesize
\begin{tabular}{ll|ll}\hline
  $2n$ & $\mathcal{NO}$  &   $2n$ & $\mathcal{NO}$ \\ \hline
 $6$ & $25=5(3+2)$   &  $54$ & $141625=5(3+2)(3+2^3)(3+2^9)$ \\
 $10$ & $45=5(3+2^2)$  & $58$ & $81935=5(3+2^{14})$ \\
 $14$ & $270=5(14+5\cdot 2^3)$ & $62$ & $26340120=5(14+5\cdot 2^5)^3$ \\
 $18$ & $275=5(3+2)(3+2^3)$ & $66$ & $982600=5(3+2)(3+2^5)^3$ \\
 $22$ & $175=5(3+2^5)$ & $70$ & $38733660$ \\
 $26$ & $335=5(3+2^6)$ & $74$ & $1310735=5(3+2^{18})$ \\
 $30$ & $16450=25(3+2^2)(14+5\cdot 2^4)$ & $78$ & $34327450=25(3+2^6)(14+5\cdot 2^{12})$\\
 $34$ & $1805=5(3+2^4)^2$            & $82$     & $5273645=5(3+2^{10})^2$  \\
 $38$ & $2575=5(3+2^9)$                & $86$     & $11240455=5(3+2^7)^3$  \\
 $42$ & $450900=25(14+5\cdot 2^3)(14+5\cdot 2^6)$ & $90$     & $25209157050$  \\
 $46$ & $51270=5(14+5\cdot 2^{11})$            & $94$     & $209715270=5(14+5\cdot 2^{23})$  \\
 $50$ & $35945=5(3+2^2)(3+2^{10})$   & $98$     & $2831158980$  \\
\hline
\end{tabular} }
\end{center}
where

\par
  $38733660=5(3+2^2)(14+5\cdot 2^3)(14+5\cdot 2^{12})$,

\par  $25209157050=5(3+2)(3+2^2)(3+2^3)(14+5\cdot 2^4)(14+5\cdot 2^{12})$,

\par
   $2831158980=5(14+5\cdot 2^3)(14+5\cdot  2^{21})$.


\section{Conclusions and further research}
\noindent
We have given an explicit representation for self-dual cyclic codes
over the finite chain ring $R=\mathbb{F}_{2^m}[u]/\langle u^k\rangle=\mathbb{F}_{2^m}
+u\mathbb{F}_{2^m}+\ldots+u^{k-1}\mathbb{F}_{2^m}$ ($u^k=0$) and a clear Mass formula to count the number
of these codes, for any integer $k\geq 2$ and positive odd integer $n$. Then, all self-dual and $2$-quasi-cyclic
codes over finite field $\mathbb{F}_{2^m}$ of length $4n$ derived from self-dual cyclic codes
over $\mathbb{F}_{2^m}
+u\mathbb{F}_{2^m}$ ($u^2=0$) of length $2n$ are determined by providing their generator matrices precisely.
Moreover,
we determine the hull
of each cyclic code
of length $2n$ over $\mathbb{F}_{2^m}+u\mathbb{F}_{2^m}$, and
give an explicit representation and enumeration for self-orthogonal cyclic codes over of length $2n$
 over $\mathbb{F}_{2^m}+u\mathbb{F}_{2^m}$.

  Giving an explicit
representation and enumeration for self-dual cyclic codes
over $R$ for arbitrary even length and considering the construction of EAQECCs
from the class of self-orthogonal (resp. orthogonal self-contained) $2$-quasi-cyclic codes of length $4n$
over $\mathbb{F}_{2^m}$ derived from self-orthogonal (resp. orthogonal self-contained) cyclic codes of length $2n$
over $\mathbb{F}_{2^m}
+u\mathbb{F}_{2^m}$ are future topics of interest.

\begin{acknowledgements}
This research is supported in part by the National
Natural Science Foundation of China (Grant Nos. 11801324, 11671235, 61971243, 61571243), the Shandong Provincial Natural Science Foundation,
China (Grant No. ZR2018BA007) and the Scientific Research Fund of Hunan
Provincial Key Laboratory of Mathematical Modeling and Analysis in
Engineering (No. 2018MMAEZD09), the Scientific Research Fund of Hubei Provincial Key Laboratory of Applied Mathematics (Hubei University) (Grant No. AM201804) and the Nankai Zhide Foundation. Part of this work was
done when Yonglin Cao was visiting Chern Institute of Mathematics, Nankai
University, Tianjin, China. Yonglin Cao would like to thank the institution
for the kind hospitality.

\end{acknowledgements}

\vskip 3mm \noindent
{\bf Appendix: All distinct self-dual cyclic codes  of
length $2n$ over the ring $R=\frac{\mathbb{F}_{2^m}[u]}{\langle u^k\rangle}$, where $3\leq k\leq 5$ and $n$ is odd}

\vskip 3mm \noindent
  By Lemma 2.3, Corollary 2.5, Theorem 3.1 and Theorem 3.3, we have the follows conclusions.

\vskip 3mm
\begin{center}
\large  $k=4$
\end{center}

\noindent
  $\diamondsuit$ \textit{Using the notation in Theorem 3.3(ii), the number of self-dual cyclic codes of length $2n$
over $\mathbb{F}_{2^m}+u\mathbb{F}_{2^m}+u^2\mathbb{F}_{2^m}+u^3\mathbb{F}_{2^m}$ $(u^4=0)$ is}
$$(1+2^m+4^m)\cdot \prod_{j=2}^\lambda(1+2^{\frac{d_j}{2}m}+2^{d_jm})\cdot\prod_{j=\lambda+1}^{\lambda+\epsilon}(9+5\cdot2^{d_jm}+4^{d_jm}).$$
\textit{Precisely, all these codes are given by}
$${\cal C}=\left(\oplus_{j=1}^\lambda \varepsilon_j(x)C_j\right)\oplus
\left(\oplus_{j=\lambda+1}^{\lambda+\epsilon}(\varepsilon_{j}(x)C_{j}\oplus\varepsilon_{j+\epsilon}(x)C_{j+\epsilon})\right),$$
\textit{where $C_j$ is an ideal of ${\cal K}_j+u{\cal K}_j+u^2{\cal K}_j+u^3{\cal K}_j$ $(u^4=0)$ listed as follows}:
\begin{description}
\vskip 2mm
\item{(i)}
 \textit{$C_1$ is one of the following $1+2^m+4^m$ ideals}:

\par
  \textit{$\langle u^2\rangle$, $\langle (x-1)\rangle$};

\par
   \textit{$\langle u^2+u(x-1)\omega\rangle$ where $\omega\in \mathbb{F}_{2^m}$ and $\omega\neq 0$};

\par
   \textit{$\langle u^2+(x-1)\omega\rangle$ where $\omega=a_0+ua_1$, $a_0,a_1\in \mathbb{F}_{2^m}$ and $a_0\neq 0$};

\par
   \textit{$\langle u^3+(x-1)\omega\rangle$ where $\omega\in \mathbb{F}_{2^m}$ and $\omega\neq 0$};

\par
  \textit{$\langle u^3,u(x-1)\rangle$}.

\vskip 2mm
\item{(ii)} \textit{Let $2\leq j\leq \lambda$. Then $C_j$ is one of the following $1+2^{\frac{d_j}{2}m}+2^{d_jm}$ ideals}:

\par
  \textit{$\langle u^2\rangle$, $\langle f_j(x)\rangle$};

\par
   \textit{$\langle u^2+uf_j(x)\omega\rangle$ where $\omega\in \Omega_{j,1}$};

\par
   \textit{$\langle u^2+f_j(x)\omega\rangle$ where $\omega=a_0(x)+ua_1(x)$,
$a_0(x)\in \Omega_{j,1}$ and $a_1(x)\in \{0\}\cup\Omega_{j,1}$};

\par
   \textit{$\langle u^3+f_j(x)\omega\rangle$ where $\omega\in \Omega_{j,1}$};

  \textit{$\langle u^3,uf_j(x)\rangle$}.

\vskip 2mm
\item{(iii)} \textit{Let $\lambda+1\leq j\leq \lambda+\epsilon$. Then the pair $(C_j,C_{j+\epsilon})$ of ideals is one of the following $9+5\cdot 2^{d_jm}+4^{d_jm}$ cases listed in the following table}:
\end{description}

{\small \begin{center}
\begin{tabular}{llll}\hline
$\mathcal{L}$ &  $C_j$ (mod $f_j(x)^2$) & $|C_j|$ & $C_{j+\epsilon}$  (mod $f_{j+\epsilon}(x)^2$)\\ \hline
$5$ & $\bullet$ $\langle u^i\rangle$ \ $(0\leq i\leq 4)$  & $4^{(4-i)d_jm}$ & $\diamond$ $\langle u^{k-i}\rangle$ \\
$4$ & $\bullet$  $\langle u^sf_j(x)\rangle$ ($0\leq s\leq 3$) & $2^{(4-s)d_jm}$ & $\diamond$ $\langle u^{4-s},f_{j+\epsilon}(x)\rangle$ \\
$3(2^{d_jm}-1)$ & $\bullet$ $\langle u^{i}+u^{i-1} f_j(x)\omega\rangle$ & $4^{(4-i)d_jm}$ & $\diamond$ $\langle u^{4-i}+u^{3-i}f_{j+\epsilon}(x)\omega^{\prime}\rangle$ \\
  & \ \ \ ($i=1,2,3$) & & \\
$4^{d_jm}-2^{d_jm}$ & $\bullet$ $\langle u^{2}+f_j(x)\vartheta\rangle$ & $4^{2d_jm}$ & $\diamond$ $\langle u^{2}+f_{j+\epsilon}(x)\vartheta^{\prime}\rangle$ \\
$2^{d_jm}-1$ & $\bullet$ $\langle u^3+f_j(x)\omega\rangle$ & $2^{4d_jm}$ & $\diamond$ $\langle u^3+f_{j+\epsilon}(x)\omega^{\prime}\rangle$ \\
$2^{d_jm}-1$ & $\bullet$ $\langle u^3+uf_j(x)\omega\rangle$ & $2^{3d_jm}$ & $\diamond$ $\langle u^2+f_{j+\epsilon}(x)\omega^{\prime},$ \\
  &  &  & \ \ \ \ $uf_{j+\epsilon}(x)\rangle$ \\
$6$ & $\bullet$ $\langle u^i,u^sf_j(x)\rangle$ & $2^{(8-(i+s))d_jm}$ & $\diamond$ $\langle u^{4-s}, u^{4-i}f_{j+\epsilon}(x)\rangle$ \\
  & \ \ \ ($0\leq s<i\leq 3$) & & \\
$2^{d_jm}-1$ & $\bullet$ $\langle u^2+f_j(x)\omega, uf_j(x)\rangle$ & $2^{5d_jm}$ & $\diamond$ $\langle u^3+uf_{j+\epsilon}(x)\omega^{\prime}\rangle$ \\
 \hline
\end{tabular}
\end{center} }

\begin{description}
\item{}
 \textit{where $\mathcal{L}$ is the number of pairs $(C_j,C_{j+\epsilon})$ in the same row, and}

\begin{description}
\item{}
  \textit{$\omega=\omega(x)\in \mathcal{F}_j=\frac{\mathbb{F}_{2^m}[x]}{\langle f_j(x)\rangle}$, $\omega\neq 0$ and}
$$\omega^{\prime}=\delta_jx^{-d_j}\omega(x^{-1}) \ ({\rm mod} \ f_{j+\epsilon}(x));$$

\item{}
  \textit{$\vartheta=a_0(x)+ua_1(x)$ with $a_0(x),a_1(x)\in \mathcal{F}_j$ and $a_0(x)\neq 0$, and}
$$\vartheta^{\prime}=\delta_jx^{-d_j}\left(a_0(x^{-1})+ua_1(x^{-1})\right)
\ ({\rm mod} \ f_{j+\epsilon}(x)).$$
\end{description}
\end{description}

\vskip 3mm
\begin{center}
\large  $k=3$
\end{center}

\noindent
  $\diamondsuit$ \textit{Using the notation in Theorem 3.3(ii), the number of self-dual cyclic codes of length $2n$
over the ring $\mathbb{F}_{2^m}+u\mathbb{F}_{2^m}+u^2\mathbb{F}_{2^m}$ $(u^3=0)$ is}
$$(1+2^m)\cdot\prod_{j=2}^\lambda(1+2^{\frac{d_j}{2}m})
\cdot\prod_{j=\lambda+1}^{\lambda+\epsilon}(7+3\cdot2^{d_jm}).$$
\textit{Precisely, all these codes are given by}
$${\cal C}=\left(\oplus_{j=1}^\lambda \varepsilon_j(x)C_j\right)\oplus
\left(\oplus_{j=\lambda+1}^{\lambda+\epsilon}(\varepsilon_{j}(x)C_{j}\oplus\varepsilon_{j+\epsilon}(x)C_{j+\epsilon})\right),$$
\textit{where $C_j$ is an ideal of ${\cal K}_j+u{\cal K}_j+u^2{\cal K}_j$ $(u^3=0)$ listed as follows}:

\begin{description}
\vskip 2mm
\item{(i)}
 \textit{$C_1$ is one of the following $1+2^m$ ideals}:

\par
  \textit{$\langle (x-1)\rangle$, $\langle u^2,u(x-1)\rangle$};
  \textit{$\langle u^2+(x-1)\omega\rangle$ where $\omega\in \mathbb{F}_{2^m}$ and $\omega\neq 0$}.

\vskip 2mm
\item{(ii)}
  \textit{Let $2\leq j\leq \lambda$. Then $C_j$ is one of the following $1+2^{\frac{d_j}{2}m}$ ideals}:

\par
  \textit{$\langle f_j(x)\rangle$, $\langle u^2, uf_j(x)\rangle$};
  \textit{$\langle u^2+f_j(x)\omega\rangle$ where $\omega\in \Omega_{j,1}$}.

\vskip 2mm
\item{(iii)}
 \textit{Let $\lambda+1\leq j\leq \lambda+\epsilon$. Then the pair $(C_j,C_{j+\epsilon})$ of ideals is one of the following $7+3\cdot 2^{d_jm}$ cases listed in the following table}:
\end{description}

\begin{center}
\begin{tabular}{llll}\hline
$\mathcal{L}$ &  $C_j$ (mod $f_j(x)^2$) & $|C_j|$ & $C_{j+\epsilon}$  (mod $f_{j+\epsilon}(x)^2$)\\ \hline
$4$ & $\bullet$ $\langle u^i\rangle$ \ $(i=0,1,2,3)$ & $4^{(3-i)d_jm}$ & $\diamond$ $\langle u^{3-i}\rangle$  \\
$3$ & $\bullet$  $\langle u^sf_j(x)\rangle$ ($s=0,1,2$) & $2^{(3-s)d_jm}$ & $\diamond$ $\langle u^{3-s},f_{j+\epsilon}(x)\rangle$ \\
$2^{d_jm}-1$ & $\bullet$ $\langle u+f_j(x)\omega\rangle$ & $4^{2d_jm}$ & $\diamond$ $\langle u^2+uf_{j+\epsilon}(x)\omega^{\prime}\rangle$ \\
$2^{d_jm}-1$ & $\bullet$ $\langle u^2+uf_j(x)\omega\rangle$ & $4^{d_jm}$ & $\diamond$ $\langle u+f_{j+\epsilon}(x)\omega^{\prime}\rangle$ \\
$2^{d_jm}-1$ & $\bullet$ $\langle u^2+f_j(x)\omega\rangle$ & $2^{3d_jm}$ & $\diamond$ $\langle u^2+f_{j+\epsilon}(x)\omega^{\prime}\rangle$ \\
$3$ & $\bullet$ $\langle u^i,u^sf_j(x)\rangle$ & $2^{(6-(i+s))d_jm}$ &$\diamond$ $\langle u^{3-s}, u^{3-i}f_{j+\epsilon}(x)\rangle$ \\
  & \ \ \ $(0\leq s<i\leq 2)$ &  & \\
 \hline
\end{tabular}
\end{center}

\begin{description}
\item{}
\textit{where $\mathcal{L}$ is the number of pairs $(C_j,C_{j+\epsilon})$ in the same row,
$\omega=\omega(x)\in \mathcal{F}_j=\frac{\mathbb{F}_{2^m}[x]}{\langle f_j(x)\rangle}$ with $\omega\neq 0$, and}
$\omega^{\prime}=\delta_jx^{-d_j}\omega(x^{-1}) \ ({\rm mod} \ f_{j+\epsilon}(x)).$
\end{description}

\vskip 3mm
\begin{center}
\large  $k=5$
\end{center}

\noindent
  $\diamondsuit$ \textit{Using the notation in Theorem 3.3(ii), the number of self-dual cyclic codes of length $2n$
over the ring $\mathbb{F}_{2^m}+u\mathbb{F}_{2^m}+u^2\mathbb{F}_{2^m}+u^3\mathbb{F}_{2^m}+u^4\mathbb{F}_{2^m}$ $(u^5=0)$ is}
$$(1+2^m+4^m)\cdot \prod_{j=2}^\lambda(1+2^{\frac{d_j}{2}m}+2^{d_jm})
\cdot\prod_{j=\lambda+1}^{\lambda+\epsilon}(11+7\cdot2^{d_jm}+3\cdot 4^{d_jm}).$$
\textit{Precisely, all these codes are given by}
$${\cal C}=\left(\oplus_{j=1}^\lambda \varepsilon_j(x)C_j\right)\oplus
\left(\oplus_{j=\lambda+1}^{\lambda+\epsilon}(\varepsilon_{j}(x)C_{j}\oplus\varepsilon_{j+\epsilon}(x)C_{j+\epsilon})\right),$$
\textit{where $C_j$ is an ideal of ${\cal K}_j+u{\cal K}_j+u^2{\cal K}_j+u^3{\cal K}_j+u^4{\cal K}_j$ $(u^5=0)$ listed as follows}:

\begin{description}
\item{(i)} \textit{$C_1$ is one of the following $1+2^m+4^m$ ideals}:

\par
  \textit{$\langle (x-1)\rangle$};

\par
  \textit{$\langle u^3+(x-1)\omega\rangle$ where $\omega=a_0+ua_1$, $a_0,a_1\in \mathbb{F}_{2^m}$ and $a_0\neq 0$};

\par
  \textit{$\langle u^4+(x-1)\omega\rangle$ where $\omega\in \mathbb{F}_{2^m}$ and $\omega\neq 0$};

\par
  \textit{$\langle u^3, u^2(x-1)\rangle$, $\langle u^4, u(x-1)\rangle$};

\par
  \textit{$\langle u^3+u(x-1)\omega, u^2(x-1)\rangle$ where $\omega\in \mathbb{F}_{2^m}$ and $\omega\neq 0$}.

\vskip 2mm
\item{(ii)} \textit{Let $2\leq j\leq \lambda$. Then $C_j$ is one of the following $1+2^{\frac{d_j}{2}m}+2^{d_jm}$ ideals}:

\par
  \textit{$\langle f_j(x)\rangle$};

\par
  \textit{$\langle u^3+f_j(x)\omega\rangle$ where $\omega=a_0(x)+ua_1(x)$, $a_0(x)\in \Omega_{j,1}$ and $a_1(x)\in \{0\}\cup\Omega_{j,1}$};

\par
  \textit{$\langle u^4+f_j(x)\omega\rangle$ where $\omega\in \Omega_{j,1}$};

\par
  \textit{$\langle u^3, u^2f_j(x)\rangle$, $\langle u^4, uf_j(x)\rangle$};

\par
  \textit{$\langle u^3+uf_j(x)\omega, u^2f_j(x)\rangle$ where $\omega\in \Omega_{j,1}$}.

\vskip 2mm
\item{(iii)}
  \textit{Let $\lambda+1\leq j\leq \lambda+\epsilon$. Then the pair $(C_j,C_{j+\epsilon})$ of ideals is one of the following $11+7\cdot 2^{d_jm}+3\cdot 4^{d_jm}$ cases listed in the following table}:
\end{description}

{\small \begin{center}
\begin{tabular}{llll}\hline
$\mathcal{L}$ &  $C_j$ (mod $f_j(x)^2$) & $|C_j|$ & $C_{j+\epsilon}$  (mod $f_{j+\epsilon}(x)^2$)\\ \hline
$6$ & $\bullet$ $\langle u^i\rangle$ \ $(0\leq i\leq 5)$ & $4^{(5-i)d_jm}$ & $\diamond$ $\langle u^{5-i}\rangle$  \\
$5$ & $\bullet$  $\langle u^sf_j(x)\rangle$  & $2^{(5-s)d_jm}$ & $\diamond$ $\langle u^{5-s},f_{j+\epsilon}(x)\rangle$ \\
  & \ \ \ ($0\leq s\leq 4$) &  & \\
$4(2^{d_jm}-1)$ & $\bullet$ $\langle u^i+u^{i-1}f_j(x)\omega\rangle$ & $4^{(5-i)d_jm}$ & $\diamond$ $\langle u^{5-i}+u^{4-i}f_{j+\epsilon}(x)\omega^{\prime}\rangle$ \\
  & \ \ \ ($i=1,2,3,4$) &  & \\
$4^{d_jm}-2^{d_jm}$ & $\bullet$ $\langle u^2+f_j(x)\vartheta\rangle$ & $4^{3d_jm}$ & $\diamond$ $\langle u^{3}+uf_{j+\epsilon}(x)\vartheta^{\prime}\rangle$ \\
$4^{d_jm}-2^{d_jm}$ & $\bullet$ $\langle u^3+uf_j(x)\vartheta\rangle$ & $4^{2d_jm}$ & $\diamond$ $\langle u^{2}+f_{j+\epsilon}(x)\vartheta^{\prime}\rangle$ \\
$4^{d_jm}-2^{d_jm}$ & $\bullet$ $\langle u^3+f_j(x)\vartheta\rangle$ & $2^{5d_jm}$ & $\diamond$ $\langle u^3+f_{j+\epsilon}(x)\vartheta^{\prime}\rangle$ \\
$2^{d_jm}-1$ & $\bullet$ $\langle u^4+f_j(x)\omega\rangle$ & $2^{5d_jm}$ & $\diamond$ $\langle u^4+f_{j+\epsilon}(x)\omega^{\prime}\rangle$ \\
$2^{d_jm}-1$ & $\bullet$ $\langle u^4+uf_j(x)\omega\rangle$ & $2^{4d_jm}$ & $\diamond$ $\langle u^3+f_{j+\epsilon}(x)\omega^{\prime},$ \\
  &  &  & \ \ \ \ $uf_{j+\epsilon}(x)\rangle$ \\
$2^{d_jm}-1$ & $\bullet$ $\langle u^4+u^2f_j(x)\omega\rangle$ & $2^{3d_jm}$ & $\diamond$ $\langle u^2+f_{j+\epsilon}(x)\omega^{\prime},$ \\
  &  &  & \ \ \ \ $uf_{j+\epsilon}(x)\rangle$ \\
$10$ & $\bullet$ $\langle u^i,u^sf_j(x)\rangle$ & $2^{(10-(i+s))d_jm}$ &$\diamond$ $\langle u^{5-s}, u^{5-i}f_{j+\epsilon}(x)\rangle$ \\
  & \ \ \ $(0\leq s<i\leq 4)$ &  & \\
$2^{d_jm}-1$ & $\bullet$ $\langle u^2+f_j(x)\omega,uf_j(x)\rangle$ & $2^{7d_jm}$ & $\diamond$ $\langle u^4+u^2f_{j+\epsilon}(x)\omega^{\prime}\rangle$ \\
$2^{d_jm}-1$ & $\bullet$ $\langle u^3+f_j(x)\omega,uf_j(x)\rangle$ & $2^{6d_jm}$ & $\diamond$ $\langle u^4+uf_{j+\epsilon}(x)\omega^{\prime}\rangle$ \\
$2^{d_jm}-1$ & $\bullet$ $\langle u^3+uf_j(x)\omega,$ & $2^{5d_jm}$ & $\diamond$ $\langle u^3+uf_{j+\epsilon}(x)\omega^{\prime},$ \\
  & \ \ \ \ $u^2f_j(x)\rangle$  &  &  \ \ \ \ $u^2f_{j+\epsilon}(x)\rangle$ \\
 \hline
\end{tabular}
\end{center} }

\begin{description}
\item{}
\textit{where $\mathcal{L}$ is the number of pairs $(C_j,C_{j+\epsilon})$ in the same row};

\begin{description}
\item{}
  \textit{$\omega=\omega(x)\in \mathcal{F}_j=\frac{\mathbb{F}_{2^m}[x]}{\langle f_j(x)\rangle}$, $\omega\neq 0$ and}
$$\omega^{\prime}=\delta_jx^{-d_j}\omega(x^{-1}) \ ({\rm mod} \ f_{j+\epsilon}(x));$$

\item{}
  \textit{$\vartheta=a_0(x)+ua_1(x)$ with $a_0(x),a_1(x)\in \mathcal{F}_j$ and $a_0(x)\neq 0$, and}
$$\vartheta^{\prime}=\delta_jx^{-d_j}\left(a_0(x^{-1})+ua_1(x^{-1})\right)
\ ({\rm mod} \ f_{j+\epsilon}(x)).$$
\end{description}

\end{description}



\end{document}